\documentclass[longauth]{aa} % for the long lists of affiliations 

\usepackage[varg]{txfonts}
\usepackage{xcolor}

\usepackage{graphicx}
\usepackage[version=4]{mhchem}
\usepackage{lscape}

\begin{document}

\title{The CARMENES search for exoplanets around M dwarfs}

\subtitle{Not-so-fine hyperfine-split vanadium lines in cool star spectra 
}

\titlerunning{Hyperfine-split vanadium lines in M dwarfs}

\authorrunning{Y.~Shan et al.}

\author{Y.~Shan \inst{\ref{inst:iag}}
\and A.~Reiners \inst{\ref{inst:iag}} 
\and D.~Fabbian \inst{\ref{inst:iag},\ref{inst:wien},\ref{inst:wien2}} 
\and E.~Marfil \inst{\ref{inst:ucm}}
\and D.~Montes \inst{\ref{inst:ucm}}
\and H.\,M.~Tabernero \inst{\ref{inst:cabesac}}
\and I.~Ribas \inst{\ref{inst:ice},\ref{inst:ieec}}
\and J.\,A.~Caballero \inst{\ref{inst:cabesac}}
\and A.~Quirrenbach \inst{\ref{inst:lsw}}
\and P.\,J.~Amado \inst{\ref{inst:iaa}}
\and J.~Aceituno \inst{\ref{inst:caha},\ref{inst:iaa}}
\and V.\,J.\,S.~B\'ejar \inst{\ref{inst:iac},\ref{inst:iac2}}
\and M.~Cort\'es-Contreras \inst{\ref{inst:cabesac}}
\and S.~Dreizler \inst{\ref{inst:iag}}
\and A.\,P.~Hatzes \inst{\ref{inst:tls}}
\and Th.~Henning \inst{\ref{inst:mpia}}
\and S.~V.~Jeffers \inst{\ref{inst:mps},\ref{inst:iag}}
\and A.~Kaminski \inst{\ref{inst:lsw}}
\and M.~K\"urster\inst{\ref{inst:mpia}}
\and M.~Lafarga\inst{\ref{inst:ice},\ref{inst:ieec},\ref{inst:war}}
\and J.\,C.~Morales\inst{\ref{inst:ice},\ref{inst:ieec}}
\and E.~Nagel \inst{\ref{inst:hs},\ref{inst:tls}}
\and E.~Pall\'e \inst{\ref{inst:iac}}
\and V.\,M.~Passegger\inst{\ref{inst:uok},\ref{inst:hs}}
\and C.~Rodriguez-L\'opez \inst{\ref{inst:iaa}}
\and A.~Schweitzer \inst{\ref{inst:hs}}
\and M.~Zechmeister \inst{\ref{inst:iag}}
}

\institute{
\label{inst:iag}Institut f\"ur Astrophysik, Georg-August-Universit\"at, Friedrich-Hund-Platz 1, 37077 G\"ottingen, Germany \\
\email{yutong.shan@uni-goettingen.de} 
\and
\label{inst:wien}University of Applied Sciences Technikum Wien, H\"ochstädtplatz 6, 1200 Wien, Austria
\and
\label{inst:wien2}Fakult\"at fur Mathematik, Universit\"at Wien, Oskar-Morgenstern-Platz 1, 1090 Wien, Austria
\and
\label{inst:ucm}Departamento de F\'{i}sica de la Tierra y Astrof\'{i}sica and IPARCOS-UCM (Instituto de F\'{i}sica de Part\'{i}culas y del Cosmos de la UCM), Facultad de Ciencias F\'{i}sicas, Universidad Complutense de Madrid, 28040, Madrid, Spain
\and
\label{inst:cabesac}Centro de Astrobiolog\'ia (CSIC-INTA), ESAC, Camino bajo del castillo s/n, 28692 Villanueva de la Ca\~nada, Madrid, Spain
\and
\label{inst:ice}Institut de Ci\`encies de l’Espai (ICE, CSIC), Campus UAB, Can Magrans s/n, 08193 Bellaterra, Spain
\and 
\label{inst:ieec}Institut d’Estudis Espacials de Catalunya (IEEC), 08034 Barcelona, Spain
\and
\label{inst:lsw}Landessternwarte, Zentrum f\"ur Astronomie der Universit\"at Heidelberg, K\"onigstuhl 12, 69117 Heidelberg, Germany
\and
\label{inst:iaa}Instituto de Astrof\'isica de Andaluc\'ia (CSIC), Glorieta de la Astronom\'ia s/n, 18008 Granada, Spain
\and
\label{inst:caha}Centro Astron\'omico Hispano-Alem\'an, Observatorio de Calar Alto, Sierra de los Filabres, 04550 G\'ergal, Spain
\and
\label{inst:iac}Instituto de Astrof\'isica de Canarias, 38205 La Laguna, Tenerife, Spain
\and 
\label{inst:iac2}Departamento de Astrof\'isica, Universidad de La Laguna, 38206 La Laguna, Tenerife, Spain
\and
\label{inst:tls}Th\"uringer Landessternwarte Tautenburg, Sternwarte 5, 07778 Tautenburg, Germany
\and
\label{inst:mpia}Max-Planck-Institut f\"ur Astronomie, K\"onigstuhl 17, 69117 Heidelberg, Germany
\and
\label{inst:mps}Max-Planck-Institut f\"ur Sonnensystemforschung, Justus-von-Liebig-Weg 3, 37077 G\"ottingen, Germany
\and
\label{inst:war}Department of Physics, University of Warwick, Gibbet Hill Road, Coventry CV4 7AL, United Kingdom
\and
\label{inst:hs}Hamburger Sternwarte, Universit\"at Hamburg, Gojenbergsweg 112, 21029 Hamburg, Germany
\and
\label{inst:uok}Homer L. Dodge Department of Physics and Astronomy, University of Oklahoma, 440 West Brooks Street, Norman, OK 73019, United States of America
}

\date{Received 11 June 2021 / Accepted 6 August 2021}

\abstract
{M-dwarf spectra are complex and notoriously difficult to model, posing challenges to understanding their photospheric properties and compositions in depth. Vanadium (V) is an iron-group element whose abundance supposedly closely tracks that of iron, but has origins that are not completely understood. } 
{Our aim is to characterize a series of neutral vanadium atomic absorption lines in the 800--910\,nm wavelength region of high signal-to-noise, high-resolution, telluric-corrected M-dwarf spectra from the CARMENES survey. Many of these lines are prominent and exhibit a distinctive broad and flat-bottom shape, which is a result of hyperfine structure (HFS). We investigate the potential and implications of these HFS split lines for abundance analysis of cool stars.
}
{With standard spectral synthesis routines, as provided by the spectroscopy software {\tt{iSpec}} and the latest atomic data (including HFS) available from the VALD3 database, we modeled these striking line profiles. We used them to measure V abundances of cool dwarfs.}
{We determined V abundances for 135 early M dwarfs (M0.0\,V to M3.5\,V) in the CARMENES guaranteed time observations sample. They exhibit a [V/Fe]-[Fe/H] trend consistent with that derived from nearby FG dwarfs. The tight ($\pm 0.1$\,dex) correlation between [V/H] and [Fe/H] suggests the potential application of V as an alternative metallicity indicator in M dwarfs. We also show hints that neglecting to model HFS could partially explain the temperature correlation in V abundance measurements observed in previous studies of samples involving dwarf stars with $T_{\rm eff} \lesssim 5300$\,K. }
{Our work suggests that HFS can impact certain absorption lines in cool photospheres more severely than in Sun-like ones. Therefore, we advocate that HFS should be carefully treated in abundance studies in stars cooler than $\sim 5000$\,K. On the other hand, strong HFS split lines in high-resolution spectra present an opportunity for precision chemical analyses of large samples of cool stars. The V-to-Fe trends exhibited by the local M dwarfs continue to challenge theoretical models of V production in the Galaxy. }


\maketitle

%\bigskip
\section{Introduction}

The analysis of detailed chemical abundances in stellar photospheres can help constrain the mechanisms of cosmic nucleosynthesis, reveal the evolution of galaxies and their stellar populations, and shed light on the formation of exoplanets \citep[e.g.,][]{Tinsley80,Timmes95,Kobayashi06,Fabbian09,Nomoto13,Kobayashi20,Fischer05,Melendez09,Adibekyan12,Dorn15,Adibekyan21}. Accurate measurements require access to spectra with high resolution and high signal-to-noise (S/N), as well as robust stellar atmosphere models and comprehensive atomic data \citep[see e.g., reviews by ][]{AllendePrieto16,Jofre19}. 

Vanadium ($_{23}$V) is an odd-$Z$ iron-peak element whose origin is not completely understood. 
It is thought to be chiefly produced in explosive silicon and oxygen burning in supernovae (SNe) of type II \citep{WW95}, with type Ia SNe supposedly also contributing to a lesser extent \citep[e.g.,][]{Clayton03,Bravo12}. 
However, when the theories are propagated through nucleosynthesis and Galactic chemical evolution (GCE) models, the predicted vanadium-to-iron ratios, [V/Fe] $\sim -0.2$ to $-0.5$\,dex, fall significantly below those observed in representative stellar populations, where [V/Fe] remains close to solar or slightly super-solar over a large range of metallicities \citep[e.g.,][]{Kobayashi11,Kobayashi20, Nomoto13,Sneden16}.    

To date, most measurements of V abundance have been performed in populations of Sun-like stars \citep[e.g.,][to name a few]{Gratton91,Feltzing98,Prochaska00b,Reddy03,BB15} or giants and subgiants \citep[e.g.,][]{Liu12,Roederer14,Hawkins16,Lomaeva19}, almost exclusively with $T_{\rm eff} \gtrsim 4000$\,K. 
The majority of studies use observations in the visual wavelength range: 400--700\,nm. 
A number of neutral and ionized V lines lie in this region, from which abundances can be constrained (and, in some cases, from as few as one single line) with either equivalent width (EW) analysis or direct spectral synthesis fits (SSF). 
While these combinations of data and methodology seem to generate robust results for FG-type stars, several studies whose samples included K dwarfs have found a large scatter in [V/Fe] and a systematic upward trend with decreasing effective temperature ($T_{\rm eff}$) below $\sim 5300$\,K \citep[e.g.,][]{Bodaghee03,Gilli06,Neves09,Adibekyan12,Tabernero12,Montes18}. 
This correlation has often been ascribed to modeling deficiencies, such as effects that are not accounted from non-local thermodynamic equilibrium (NLTE). 

Incidentally, $^{51}_{23}$V, the most abundant isotope accounting for $>99\,\%$ of naturally occurring vanadium, has a rather high nuclear spin (+7/2) and shows pronounced hyperfine structure (HFS), which are shifts in atomic energy levels. These shifts split atomic transitions into multiple, usually closely spaced components. 
The manifestation of HFS in a number of absorption lines, including those of V, was already noticed in solar spectra by \citet{Abt52}. 
However, due to limitations in atomic data, and perhaps also in awareness, not all stellar population studies to date have completely accounted for HFS in their spectral modeling. 
Interestingly, all the aforementioned studies of K dwarfs that have observed a correlation between V abundance and $T_{\rm eff}$ neglected to include HFS components in their line lists. 

A number of authors have cautioned against inappropriate modeling of HFS in stellar abundance analysis, either in terms of the use of inaccurate HFS atomic data or a complete absence of consideration for HFS. 
For example, \citet{Prochaska00a} and \citet{Jofre15b} found notable abundance systematics for certain elements (including V) arising from the suboptimal treatment of HFS. 
When presenting the Hypatia Catalog (a large compilation of detailed elemental abundance measurements for stars in the solar neighborhood), \citet{Hinkel14} also commented on the increased scatter seen in literature abundances in elements such as Mn and V, speculating that a ``casual'' treatment of HFS could be to blame. 
On the other hand, several investigations have reached the broad conclusion that abundance measurements tend to be relatively insensitive to the inclusion of HFS \citep[e.g.,][]{delPeloso05, Takeda07, Hinkle16, Jofre17}, which may appear to contradict the former set of studies. 
However, these latter works were mostly based on the analysis of a handful of elements in FG-type dwarfs over limited wavelength ranges, and their conclusions chiefly apply to the final results, which are averages over particular ensembles of chosen lines. For specific individual lines and, especially, for cooler stars, the latter studies have also noted significant HFS-related deviations. 

Until very recently, elemental abundance studies have largely shunned M-type stars, as their cool temperatures promote the presence of molecules, giving rise to forests of poorly modeled lines that veil the continuum and distort atomic features.
Elemental abundances other than iron have been characterized for only a handful of M dwarfs \citep{Veyette17, Souto17, Souto18, Abia20, Ishikawa20}, mostly from high-resolution spectra in the $YJH$ bands in the near-infrared (NIR) region ($\lambda \gtrsim$ 960 -- 1700\,nm). 
Unfortunately, in M dwarfs, the $YJH$ bands host very few useful V lines. 
As a result, V abundance has been quantified for only one M dwarf, namely \object{Kepler-138} (M1\,V, $T_{\rm eff} \sim 3800$\,K), using one line in the APOGEE $H$ band ($\lambda \sim 1592.4$\,nm; \citealt{Souto17}).
More recently, \citet{Maldonado20} determined abundances for a number of elements, including V, for $\sim200$ M dwarfs. 
Their approach was not based on the study of individual line profiles, but rather on broad features in severely degraded HARPS spectra ($\sim$ 400--700\,nm, $\mathcal{R} \sim$ 115\,000 to $\mathcal{R} \sim$ 1000--2000). The features that best describe the degraded spectra were identified using principal component analysis, then correlated to abundances using a calibration sample of 19 FGK+M binaries. 
While their novel technique achieved impressive success for many of the elements considered, the V results presented in \citet{Maldonado20} were dubious and significantly deviated from patterns measured for ensembles of FGK stars. 
Therefore, it appears that no robust analysis of vanadium on a large sample of M dwarfs has been performed yet. 

\begin{figure*}
    \centering
    \includegraphics[width=0.95\textwidth]{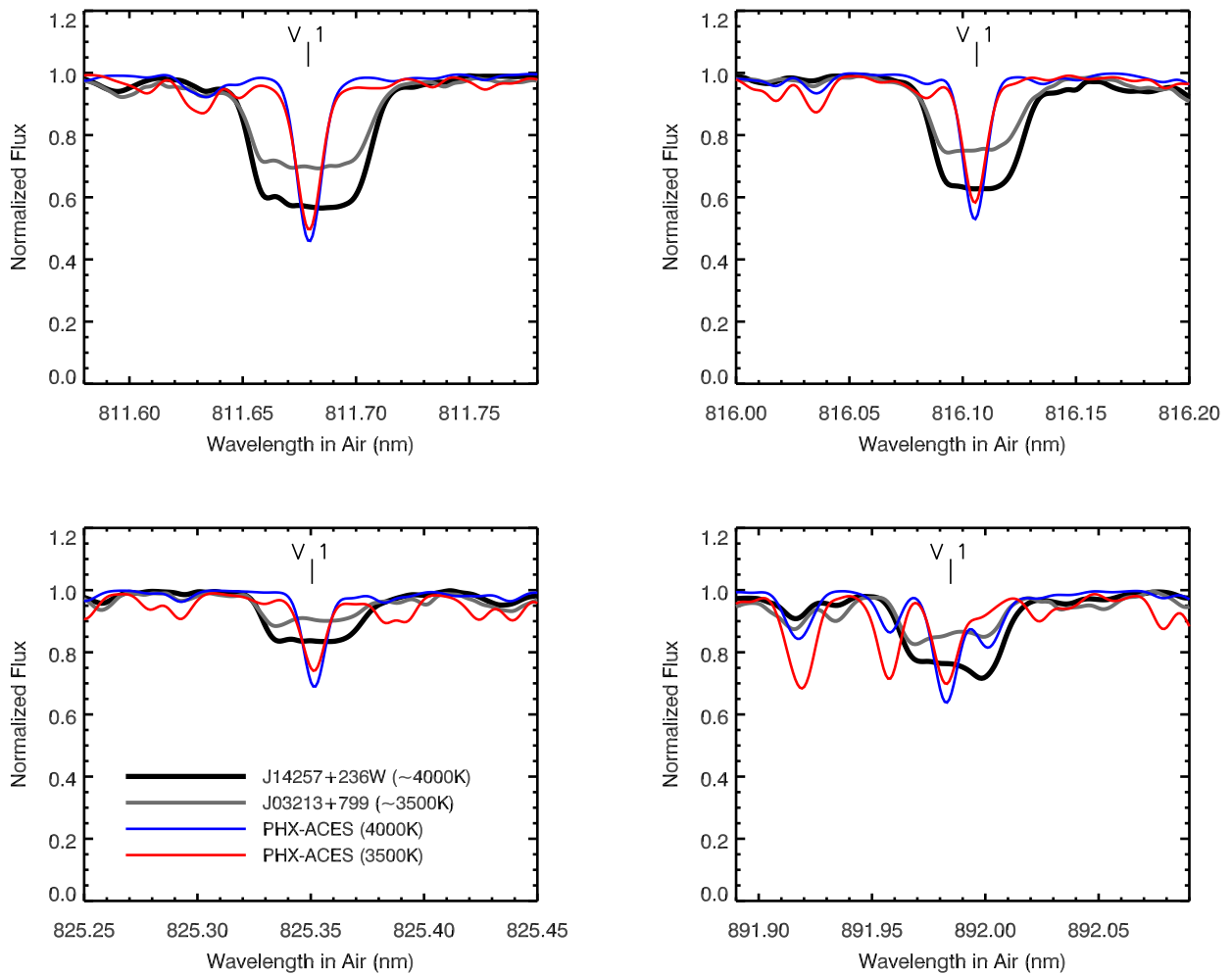}
    \caption{Representative vanadium features in normalized, telluric absorption corrected, and co-added CARMENES spectra for M0.0\,V (black, J14257+236W) and M2.0\,V (gray, J03213+799) stars. Overplotted are PHOENIX-ACES spectra (1D LTE) convolved to the CARMENES instrumental resolution ($\mathcal{R}$ = 94\,600) for 4000\,K (blue) and 3500\,K (red) dwarfs for comparison. Wavelengths are given in terms of the rest frame as measured in air.}
    \label{f:representative-vlines}
\end{figure*}

Meanwhile, the CARMENES survey has provided high-resolution, high-quality spectra for hundreds of nearby M dwarfs. 
Some of these data were recently used to study rubidium, strontium, and zirconium in 57 local M dwarfs \citep{Abia20}. 
In this present work, we focus on vanadium. In the wavelength region of 800\,nm to 910\,nm, which hosts a relatively low density of molecular lines, there exist about a dozen features of V~{\sc i}, many of them being extremely strong and prominently split by HFS. 
In some cases, the line shape deviations in these high-resolution spectra are so large that they are very far from being degenerate with any other conventional broadening parameters. We demonstrate that the striking vanadium line profiles seen in the CARMENES spectra can be excellently reproduced using the latest available atomic data and standard stellar models, assuming reasonable stellar parameters. From these features, we extracted robust vanadium abundances for 135 early M dwarfs and show that they track iron with little scatter, matching the patterns seen for local FG-type dwarfs.
         
This paper is organized as follows.
Section~\ref{ss:carm} describes the CARMENES survey and the spectroscopic data products used for this work. 
Section~\ref{ss:square-buckets} offers a heuristic description of the HFS-split vanadium features. 
We give an overview of HFS and its role in spectral modeling and abundance studies in Section~\ref{s:hfs}. 
In Section~\ref{s:v-abunds}, we document our effort to determine vanadium abundances for a subsample of early M dwarfs from the CARMENES survey. 
A comparison with the literature and discussion can be found in Section~\ref{s:discussion}. 
In particular, the apparent overabundance of vanadium measured in some studies of K dwarfs is explored in Section~\ref{ss:k-dwarfs}, and Section~\ref{ss:metallicity-tracer} briefly examines the potential of vanadium abundance as a metallicity indicator for late-type stars. 
We give our conclusions in Section~\ref{s:conclusion}.

%==================================================================
\section{Vanadium lines in CARMENES spectra}\label{s:carmenes-spectra}

\subsection{The CARMENES survey}\label{ss:carm}

CARMENES is a spectroscopic survey of nearby M dwarfs to search for planets using the radial velocity method. The observations are conducted with the 3.5\,m telescope at the Calar Alto Observatory in Almer\'ia, Spain, using the homonymous state-of-the-art fiber-fed high-resolution spectrograph with optical (VIS: $\Delta \lambda \approx$ 520--960\,nm, $\mathcal{R} \sim 94\,600$) and near-infrared (NIR: $\Delta \lambda \approx$ 960--1710\,nm, $\mathcal{R} \sim 80\,400$) channels \citep{Quirrenbach14,Quirrenbach18}. 
Since January 2016, the CARMENES project has collected high-quality spectra over multiple epochs of $\sim 380$ stars across the full M-dwarf range as a part of its guaranteed time observations (GTO) program \citep[][hereafter Schw19]{Reiners18, Schweitzer19}. 
The utility of this data set has been demonstrated in numerous planet discoveries and atmospheric characterizations, as well as studies of stellar properties, such as activity and fundamental parameters \citep[see][for an overview]{Quirrenbach20} 

As multiple spectra are taken over time for each target, they can be combined into a single ``template spectrum'' optimized for detailed stellar characterization, as follows: the raw spectra have been processed with the CARMENES pipeline {\tt caracal}, which is based on flat-relative optimal extraction \citep{Zechmeister14} and wavelength calibration, which utilizes spectra from hollow cathode lamps and Fabry-P\'erot etalons \citep{Bauer15}. 
The spectra and, thus, the templates are corrected for telluric absorption following the template division telluric modeling approach \citep{Nagel19}, which employs {\tt molecfit}\footnote{\url{http://www.eso.org/sci/software/pipelines/skytools/molecfit}} \citep{Smette15}. In a further iteration, the {\tt serval}\footnote{\url{https://github.com/mzechmeister/serval}} code \citep{Zechmeister18} coadds all spectra of each star via a B-spline regression and creates a telluric-free high S/N template spectrum \citep{Caballero16}.  
Each template spectrum, covering from 520\,nm to 1710\,nm with a few inter-order gaps mostly in the $H$ band, comprises on average $\sim50$ individual spectra, and reaches S/N of $> 100$ virtually throughout the entire wavelength range. 
The high resolution, high S/N, and virtual absence of telluric lines make these template spectra suitable for identifying and modeling fine features intrinsic to a given stellar spectrum.

\subsection{Heuristic description of ``square-bucket'' features associated with V~{\sc i} lines}\label{ss:square-buckets}

Compared to shorter wavelengths, the 800--910\,nm region of M dwarf spectra is noticeably less impacted by molecular absorption in the stellar photosphere. 
After correction for telluric absorption, a series of features stand out in this wavelength range in the template spectra of many CARMENES GTO stars (see Fig.~\ref{f:representative-vlines} for examples, and Figs.~\ref{f:4000k_representative_spectrum} and~\ref{f:3500k_representative_spectrum} for the whole 800--910\,nm region). 
These features are striking in appearance owing to their strength and breadth, and in particular their well-defined sharp corners, which cannot be reconciled with conventional broadening mechanisms of a single absorption line. Rather, each ``bucket'' resembles the smearing out of a forest of narrow, adjacent components. Some of these components are even marginally resolved into ``wiggles''. These absorption profiles are ostensibly mismatched with synthetic spectra found in standard libraries (e.g., PHOENIX-ACES, \citealt{Husser13}). Many of these features persist throughout the early M-dwarf range. They are associated with transitions of neutral vanadium atom. 

\begin{figure}
    \centering
    \hspace{-1cm}
    \includegraphics[width=0.5\textwidth]{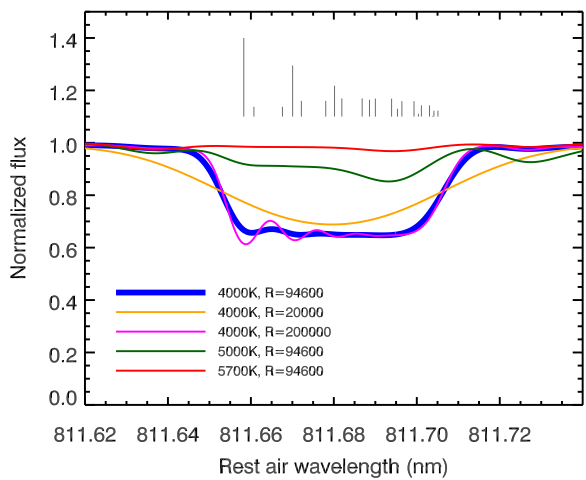}
    \caption{Synthetic spectra around the dramatically split V~{\sc i} $\lambda 811.7$\,nm synthesized by {\tt{iSpec}} using the latest VALD3-HFS line list for solar vanadium abundance. This feature comprises 21 components, indicated by gray vertical lines whose heights scale with their expected relative strengths at $T = 4000$\,K, with EW used as a proxy for strength, $\log({\rm EW}) = \log(gf\lambda) - \theta\chi$, where $\lambda$ is the transition wavelength in \AA, $\theta = 5040/T$, temperature $T$ is in K, and $\chi$ is the lower excitation potential of the transition in eV. The strongest components toward the left side of the feature manifest themselves as marginally resolved ``wiggles'' in spectra with CARMENES-like resolution.  }
    \label{f:lowres-vlines}
\end{figure}

The fact that these particular V~{\sc i} transitions are affected by such substantial splitting in cool dwarfs may have received little attention because no previous program has conducted a dedicated M-dwarf survey at high resolution, with high S/N, and at multiple epochs so as to enable robust telluric corrections in the red optical wavelength range, as CARMENES has.
The specific lines belonging to the V~{\sc i} transitions that we consider here diminish in strength toward higher temperatures, nearly vanishing at about $5000$\,K, as most neutral vanadium becomes ionized (we note that V~{\sc ii} transitions also exist, but are extremely weak in M dwarfs in the optical wavelength range). These lines also lose their ``corners'' in lower resolution ($\mathcal{R} \sim 20\,000$) spectra (see Fig.~\ref{f:lowres-vlines}). In cool dwarfs, many V~{\sc i} transitions are also visible at shorter wavelengths (e.g., 500--700\,nm), but generally appear less broadened and significantly more blended with forests of molecular absorption. 

%==================================================================
\section{Modeling the broad vanadium features with hyperfine structure}\label{s:hfs}

\subsection{Hyperfine splitting in stellar abundance studies}\label{ss:hfs-lit}

The broad V~{\sc i} features are a consequence of HFS splitting, a well-understood quantum effect arising from the interaction between electron angular momentum and nuclear magnetic moment, resulting in (typically) hyper-finely split energy levels within an atom \citep[e.g.,][]{CondonShortley63,Sobelman06}. 
HFS manifests itself in a number of atomic transitions, most notably in odd-$Z$ elements where an unpaired proton gives rise to a nonzero nuclear spin that, together with a large nuclear $g$ factor, can result in a large nuclear magnetic moment. 
Besides V, other examples of prominent elements of astrophysical interest whose most common isotopes exhibit a high nuclear magnetic dipole moment include Mn, Co, Cu, Sc, Eu, and Rb. 
For these elements, failure to account for HFS could result in inadequately modeled line profiles and systematically incorrect abundances.

The importance of accurate HFS modeling in spectral analysis has gained recognition through works such as that of \citet{Prochaska00a}, who pointed out that incorrect HFS data for Mn and Sc had led to claims of spurious trends between abundances and metallicity and inappropriate conclusions about the mechanisms of Galactic nucleosynthesis \citep[see also, e.g.,][]{Booth83,Vitas03,Bergemann10,North12,Jofre19}. HFS is now routinely incorporated in stellar abundance determinations of many elements. 
However, it has also been neglected in other notable studies \citep[e.g.,][but see \citealt{Adibekyan16b} which does account for HFS]{Feltzing98,Neves09,Adibekyan12}. One reason for the lack of universal consideration is that, for many lines, the abundance biases introduced by neglecting HFS are minuscule because the shape distortions due to HFS are subtle and, for weak lines in the unsaturated regime, HFS preserves the equivalent widths. Another important obstacle in the way of accounting for HFS effects in spectral modeling might simply be the incompleteness or lack of accessibility to atomic data.
The basic physics measurements required to characterize HFS patterns are not available for many split transitions and, even for those energy levels that have been measured, it is nontrivial to convert raw laboratory data into line information directly readable by standard software for atmosphere synthesis\footnote{For a transition affected by HFS, laboratory data, if existent, are often given in terms of the magnetic dipole, $A$, and electric quadrupole, $B$, constants for the unsplit energy levels involved. Together with knowledge of the angular momentum quantum number, $J$, and nuclear spin, $I$, the HFS line component parameters can be computed via the Casimir formula and the H\"{o}nl-Kronig intensity rule (prescriptions described in, e.g., \citealt{Woodgate80,McWilliam13,Pakhomov19}).}. 
On the other hand, databases that do include HFS information tend to supply the HFS components in files separate from their default line lists \citep[e.g.,][]{Kurucz11}\footnote{\url{ http://kurucz.harvard.edu/linelists.html}}. This could hinder widespread usage, since it gives HFS an air of exoticism and requires extra initiative from the user to incorporate them into their studies.

The precise impact of unmodeled HFS on abundance measurements is correlated with line strength and depends strongly on the spectral type, the element, and the specific set of lines used in the analysis. However, the exact nature of these dependences is difficult to characterize, as the split patterns appearing for each transition are not simple functions of wavelength, temperature, or composite strength, and must be determined on a case-by-case basis. Several authors have investigated the magnitude of systematic errors introduced by neglecting HFS in FG- (and occasionally K-) type stars in the visual wavelength range for various elements, using small sets of lines \citep[e.g.,][]{delPeloso05,Takeda07,Jofre15b,Jofre17,Hinkle16}. At least for the FG-type stars, the biases are often within typical uncertainties, especially when an abundance measurement is based on the mean over several lines. However, significant deviations were detected in specific individual lines, especially in strong lines toward the cool ends of the considered stellar samples. Empirically, the deviations were particularly evident in the works by \citet{Jofre15b} and \citet{Jofre17}, where stars cooler than $T_{\rm eff} \lesssim 5000$\,K like Arcturus were also studied. Therefore, to obtain accurate elemental abundance measurements for arbitrary types of stars, it is important to properly model all the HFS-affected lines whenever possible.

\subsection{Using HFS components for V~{\sc i} from the latest update in VALD3}\label{ss:hfs}

\begin{figure*}
    \centering
    \includegraphics[width=0.95\textwidth]{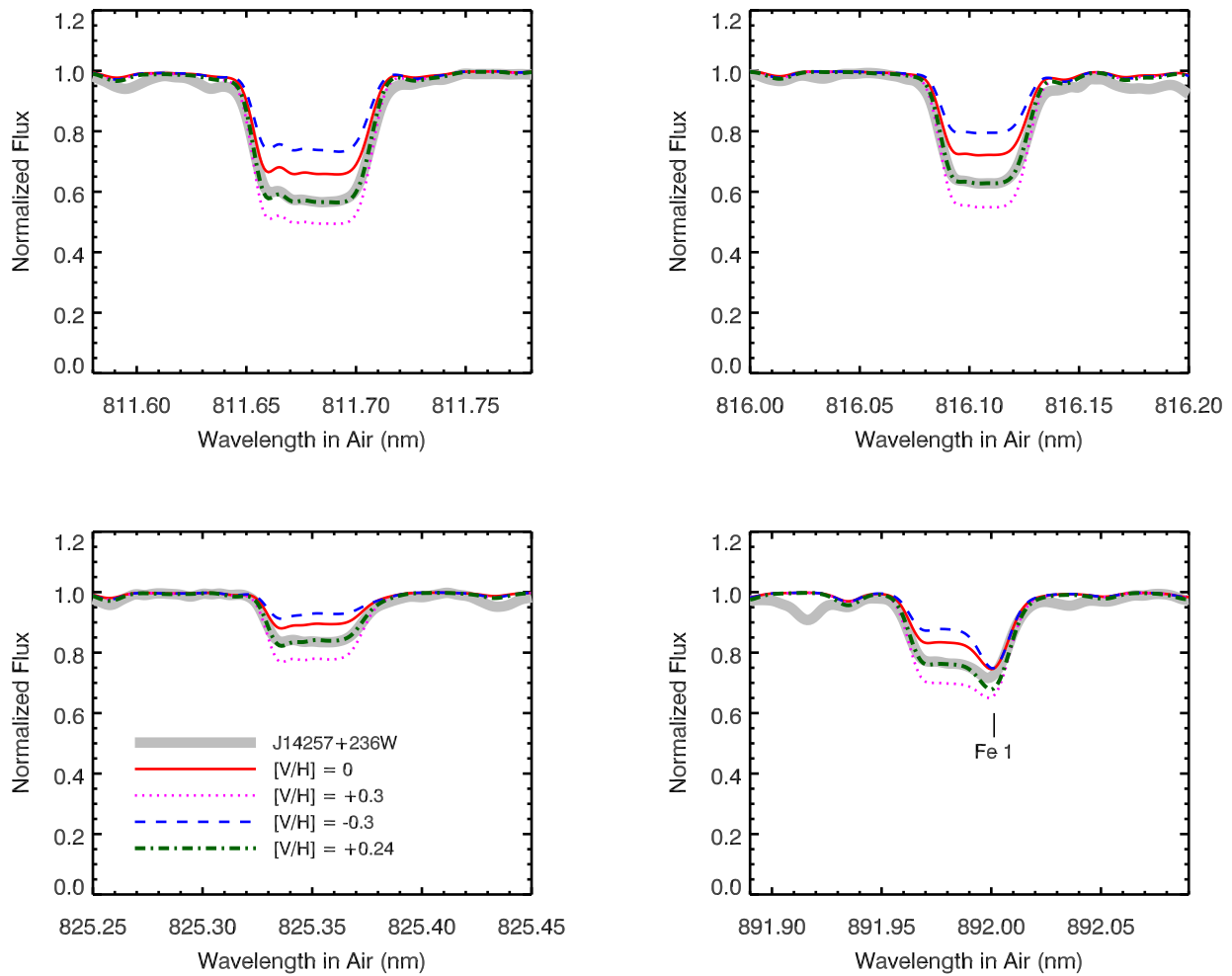}
    \caption{Illustration of synthetic spectra generated with {\tt iSpec} from the new VALD3 HFS atomic data for V compared with the CARMENES telluric-corrected composite spectrum of J14257+236W (gray) in several hyperfine-split V lines. 
    Red solid, blue dashed, and magenta dotted lines are generic models with $T_{\rm eff} = 4000$\,K, $\log{g}$ = 4.5, [M/H] = 0.0, and V abundances of solar (i.e., $A_\odot({\rm V}) = \log (N_{\rm V}/N_{\rm H})_\odot = -8.11$, \citealt{Asplund09}), +0.3 dex, and --0.3 dex. 
    The best-matching model in green dash-dotted lines takes the basic stellar parameters from Schw19: $T_{\rm eff} = 4024$\,K, $\log{g}$ = 4.64, [M/H] = +0.17, and $A({\rm V})$ = --7.87 (i.e., [V/H] = +0.24), which simultaneously reproduces the overall shape of all the features. 
    We note that the feature at 891.98\,nm is blended with an Fe~{\sc i} line at 892.0\,nm.}
    \label{f:J14257+236W_variable-AV}
\end{figure*}

HFS data have recently been added to the Vienna Atomic Line Database (VALD3), as detailed by \citet{Pakhomov17,Pakhomov19}. This supplementary database appears to be the result of an extensive effort to gather the latest atomic data from the literature, compute the associated HFS patterns, and output them into user-friendly forms. As of May 2020, these data have been readily accessible in the familiar format via the standard VALD3 web interface\footnote{\url{http://vald.astro.uu.se}} by selecting the ``HFS splitting'' option. The atomic data collection for vanadium therein includes, among others, recent results from the Wisconsin atomic group \citep{Lawler14,denHartog14,Wood14}. 

We downloaded and converted these line lists into a format recognized by the stellar spectroscopy software package {\tt{iSpec}} \citep{BC14,BC19}. Using {\tt{iSpec}}, we simulated a number of broad vanadium features between 800\,nm and 910\,nm, seen in the CARMENES VIS spectra. With the settings described in Sect.~\ref{ss:ssf-specs} and~\ref{ss:carm-gto-v}, we generated model spectra that satisfactorily reproduce the observed prominent V~{\sc i} line profiles in early M dwarfs, an example of which is shown in Fig.~\ref{f:J14257+236W_variable-AV}. The depths of the features appear to exhibit a high sensitivity to the exact V abundance value assumed. By tweaking the V abundance in the synthetic spectra, excellent matches can be made simultaneously by the model to all the features that we considered, suggesting the potential effectiveness of these features for abundance determination using the spectral synthesis technique.

%==================================================================
\section{Measuring vanadium abundances of cool stars}\label{s:v-abunds}

The presence of strong and modelable vanadium features in a relatively clean region of cool star spectra points to an opportunity to characterize V abundances of all stars of similar type. We used spectral synthesis fits to derive vanadium abundances for a large number of early M dwarfs in the CARMENES GTO sample.

\subsection{Atomic data}\label{ss:line-lists}

We used the ``extract stellar'' option of the VALD3 web interface to download a list of all line data, including HFS-split-line components, in the vicinity of the regions defined in Table~\ref{t:vline-regs} with theoretical depth exceeding $10^{-3}$ for 4000\,K and 3500\,K dwarf stars. 
The 4000\,K line list was used for stars with $T_{\rm eff} \gtrsim 3700$\,K and the 3500\,K line list was for those with $T_{\rm eff} \lesssim 3700$\,K. These line lists include all atomic species, as well as several molecules including TiO, MgH, CH, CO, \ce{C2}, OH, and CN.      

\subsection{Line region selection}\label{ss:line-regs}

Through systematic visual inspection of V~{\sc i} lines in the M-dwarf spectra at 800--910\,nm, we identified 13 regions that are well modeled using the new VALD3 atomic line lists for stars around 4000\,K. For stars with $T_{\rm eff}$ closer to 3500\,K, where molecular contamination becomes more significant, there are nine such regions for which the profile match quality is tolerable. 
Therefore, in choosing the set of lines to fit, we made a distinction between earlier- ($T_{\rm eff} \gtrsim 3700$\,K) and later- ($T_{\rm eff} \lesssim 3700$\,K) type M dwarfs. Overall, our selected line regions have a variety of strengths and breadths attributed to HFS, and are potentially suitable for V abundance determination. These regions and the number of split components that compose them are listed in Table~\ref{t:vline-regs} and illustrated in Figs.~\ref{f:4000k_fitted_lines} and~\ref{f:3500k_fitted_lines}. We note that the line regions with fewer split components tend to look less ``square''.  

\begin{table}[]
\begin{center}
\caption{Wavelength regions used for the V abundance fits.}
\label{t:vline-regs}
\begin{tabular}{cccccc}
\hline\hline
\noalign{\smallskip}
$\lambda_{\rm central}$  & $\lambda_{\rm low}$ & $\lambda_{\rm upp}$ & \# comps. & $T_{\rm eff} \gtrsim$ & $T_{\rm eff} \lesssim$ \\
(nm) & (nm) & (nm) & modeled\textsuperscript{a} & 3700\,K & 3700\,K \\
\noalign{\smallskip}
\hline
\noalign{\smallskip}
  802.7366 &    802.70 &    802.90 & 18 & yes & yes \\
  809.3468 &    809.32 &    809.37 & 12 & yes & yes \\
  811.6789 &    811.64 &    811.72 & 21 & yes & yes \\
  814.4560 &    814.44 &    814.47 & 6 & yes & yes \\
  816.1062 &    816.06 &    816.15 & 16 & yes & yes \\
  818.6728 &    818.63 &    818.70 & 10 & yes &  no \\
  824.1599 &    824.13 &    824.17 & 6 & yes & yes \\
  825.3506 &    825.30 &    825.39 & 18 & yes & yes \\
  825.5896 &    825.56 &    825.61 & 12 & yes & yes \\
  891.9847 &    891.95 &    892.02 & 18 & yes & yes \\
  893.2947 &    893.26 &    893.34 & 12 & yes &  no \\
  903.7613 &    903.74 &    903.79 & 4 & yes &  no \\
  904.6693 &    904.64 &    904.70 & 10 & yes &  no \\
\noalign{\smallskip}
\hline
\end{tabular}
\end{center}
Notes: 
\textsuperscript{a}: Number of HFS components included in the line model.
\\
\end{table}

\subsection{Synthetic spectra and fitting algorithm}\label{ss:ssf-specs}

As described in Section \ref{ss:hfs}, we used the {\tt{iSpec}} stellar spectroscopy software to synthesize and fit model spectra generated from the new line lists. {\tt{iSpec}} is a highly versatile program developed for stellar parameter and abundance analysis. It allows users to mix and match suites of atmosphere structure grids and radiative transfer codes \citep{BC19}, and comes preloaded with several popular options of both. The user provides the program with a solar elemental abundance pattern and line lists in the wavelength regions under consideration. Then, given a set of stellar properties ($T_{\rm eff}$, $\log g$, [M/H], $\alpha$), broadening parameters ($v_{\rm mic}$, $v_{\rm mac}$, $v\sin i$, instrumental resolution) and individual abundances, {\tt{iSpec}} can synthesize the stellar spectra by interpolating the atmosphere structure grid. It can also perform the inverse problem, deriving fundamental parameters and abundances from an input spectrum and line region definitions, using either SSF (based on an iterative $\chi^2$ minimization algorithm) or EW. The continuum model can be found by fitting splines to a version of the spectra smoothed with maximum and median filters, as described by \citet{BC14} and \citet{BC19}. The spline and window parameters are specified by the user. For broadening, {\tt{iSpec}} uses a radial-tangential kernel for macroturbulence and the standard Gray kernel for rotation \citep{Gray92}. For comparison with observed spectra, the synthesized spectra are convolved with a Gaussian line profile corresponding to the user-specified resolution of instrument.     
For all of our analyses, we chose the MARCS 1D, plane-parallel, LTE atmosphere models \citep{Gustafsson08} paired with the radiative transfer code {\tt Turbospectrum} version 15.1 \citep{Plez12}, both as provided in {\tt{iSpec}}. Our solar abundance pattern followed \citet{Asplund09}. We used the {\tt{iSpec}} SSF scripts to perform line-by-line fits to the applicable V~{\sc i} line regions, varying only V abundance and fixing all other fit parameters to those externally determined or assumed. Where only [Fe/H] is available (e.g., in the literature), we take it as a proxy for [M/H], {\tt{iSpec}}'s scaling factor for the overall elemental abundance pattern
([M/H] is a measure of the average overall metal content, which is expected to deviate from [Fe/H], the iron abundance, at low metallicities, mainly due to an enhancement of $\alpha$-elements. However, the distinction between [Fe/H] and [M/H] is sometimes casually treated in the M dwarf literature, meaning that some published [Fe/H] values may be closer to [M/H] -- see \citealt{Marfil21} for a relevant discussion).

\subsection{Test case: Arcturus}\label{ss:arcturus}

Arcturus is a K giant with publicly-accessible, high-resolution ($\mathcal{R} \sim 150\,000$), high signal-to-noise (S/N $\sim 1000$) {\'e}chelle spectra across the entire optical wavelength range \citep[373--930\,nm,][]{Hinkle00}. It is also the only late-type star with V abundance exquisitely measured from 55 lines at optical wavelengths (550--910\,nm) with a complete accounting of the most up-to-date knowledge of the atomic parameters and HFS effects \citep{Wood18}. This makes Arcturus a suitable benchmark target on which to test our method. Our independent selection of line regions turns out to be very similar to those used by \citet{Wood18} in the 800--910\,nm range. 
We excluded the V~{\sc i} $\lambda$897.1673\,nm and $\lambda$908.5231\,nm lines because they appear to be affected by relatively heavy molecular blending in early M-dwarf spectra.     

Where possible, we adopted identical stellar parameters as used in the analysis of \citealt{Wood18}: $T_{\rm eff} = 4275$\,K, $\log{g} = 1.3$, $v_{\rm mic} = 1.6$\,km\,s$^{-1}$, and [Fe/H] = $-0.55$\,dex, taken from \citet{Peterson17}. 
In addition, we fixed $v\sin i = 1.5$\,km\,s$^{-1}$ and $v_{\rm mac} = 5.2$\,km\,s$^{-1}$, as measured by \citet{GrayBrown06} for Arcturus. 
The fitted synthetic spectra were convolved to $\mathcal{R} = 150\,000$ using a Gaussian profile to match the resolution of the Arcturus spectrum. We used {\tt{iSpec}}'s default continuum settings whereby quadratic B-splines were fitted to 5\,nm segments filtered using the ``median+max'' scheme, where the median filter was 0.05\,nm and the maximum filter was 1.0\,nm wide. 
We performed line-by-line fits to the 13 lines listed in Table~\ref{t:vline-regs}, arriving at $\langle\log \epsilon({\rm V})\rangle = 3.55 \pm 0.01$ (sample standard deviation $\sigma = 0.04$)\footnote{$\log \epsilon({\rm X}) \equiv \log_{10} (N_{\rm X}/N_{\rm H}) + 12$.}. 
Our value compares very well with the determination by \citet{Wood18} of $\langle\log \epsilon({\rm V})\rangle = 3.54 \pm 0.01$ from 55 lines ($\sigma = 0.04$).

\subsection{CARMENES sample selection}\label{ss:subsample}

We visually inspected the telluric-corrected template spectra of 331 CARMENES GTO targets to identify stars that prominently exhibit the V HFS features of interest and in which these features can be adequately reproduced by the models. We observed that spectra of M dwarfs with $T_{\rm eff} \lesssim 3400$\,K tend to be very contaminated in a majority of the lines and surrounding continuum regions that we considered (from Section~\ref{ss:line-regs}), whereas those with $T_{\rm eff} \gtrsim 3500$\,K and $v\sin i < 4$\,km\,s$^{-1}$ are well-fit in most of the line regions under consideration. 
Based on the $T_{\rm eff}$ determined for this sample from Schw19 whenever possible and, otherwise, from \citet{Passegger19} and \citet{Passegger18}, we removed 117 stars with $T_{\rm eff} < 3400$\,K. 
A further 34 stars had no measured $T_{\rm eff}$ in any of the references given above because they are fast rotators, are very late-type stars, or have an S/N that is too low. Therefore, we discarded those as well. 
Of the 117 stars with $T_{\rm eff} > 3500$\,K, we retained all except J16102--193, which has a $v \sin i = 7.3$\,km\,s$^{-1}$ \citep{Reiners18}, J11201--104 and J18174+483, which show large broadening in the V features in excess to that accountable by the measured $v\sin i$, J00162+198W and J05532+242, which are double-lined spectroscopic binaries \citep{Baroch18}, J15474--108, which is a spectroscopic triple \citep{Baroch21}, and J04219+213, which exhibits large spurious features. 
This left 110 stars. In the remaining sample of 63 stars with 3400\,K $\leq T_{\rm eff} \leq$ 3500\,K we first eliminated four stars that have a measured $v\sin i > 4$\,km\,s$^{-1}$, leaving 59 stars. We assessed these by eye and picked out a further 25 stars where a majority of the features appear to be fit decently with the model spectra. The resulting sample consisted of $110 + 25 = 135$ early M dwarfs, all of which have $T_{\rm eff} \gtrsim 3400$\,K and $v\sin i < 4$\,km\,s$^{-1}$.

\subsection{Vanadium abundances for the CARMENES GTO subsample}\label{ss:carm-gto-v} 

We fit to the V regions as listed in Table~\ref{t:vline-regs} for this subsample of early M dwarfs from the CARMENES GTO program\footnote{The following V~{\sc i} lines were excluded from the fit for the following objects: 
J12248--182: $\lambda$825.3506\,nm (too weak); 
J14082+805: $\lambda$816.1062\,nm (contaminated); 
J16167+672S: $\lambda$818.6728\,nm and $\lambda$825.3506\,nm (contaminated); 
J17355+616: $\lambda$818.6728\,nm (contaminated).}. 
In these fits, the stellar parameters $T_{\rm eff}$, $\log g$, and [Fe/H] were fixed to those derived by Schw19 based on PHOENIX-ACES model fits to metal lines in the same spectra. 
{\tt{iSpec}} picks an atmosphere structure model consistent with the expected $\alpha$ enhancement as a function of [M/H] through the usual relation ([$\alpha$/Fe] = 0.0 for [M/H] = 0.0, increasing to +0.4 for [M/H] = --1.0; \citealt{Gustafsson08}). 
Since macroturbulence velocities, $v_{\rm mac}$, are not well-characterized for cool dwarfs, but are likely to be small (e.g., extrapolation from Fig.~17.10 in \citealt{Gray92} extends below 1\,km\,s$^{-1}$ for stars later than K3\,V -- see also \citealt{Lindgren16}), we set the $v_{\rm mac}$ parameter to 1.0\,km\,s$^{-1}$ in our fits.
The microturbulence velocity $v_{\rm mic}$ is determined by {\tt{iSpec}}'s built-in function and depends on $T_{\rm eff}$, $\log g$, and [M/H]. For M dwarfs, $v_{\rm mic}$ as assigned by {\tt{iSpec}} is generally between 0.8\,km\,s$^{-1}$ and 0.9\,km\,s$^{-1}$. 
The values recommended by \citet{Husser13} based on convective velocities from the PHOENIX-ACES models are approximately $v_{\rm mic} \sim 0.5$\,km\,s$^{-1}$. 
We checked that the final fit results are insensitive to such small $v_{\rm mic}$ variations. 
Where possible, we used the $v\sin i$ values measured by \citet{Reiners18} for this sample. When only an upper limit of 2\,km\,s$^{-1}$ is available, we set $v\sin i$ = 2\,km\,s$^{-1}$. 
We convolved the synthesized spectra with a Gaussian instrumental line profile corresponding to the resolving power of the CARMENES VIS spectrograph. As in the Arcturus fit, the local continuum normalization was done using the ``median+max'' scheme with window length of 0.05\,nm and 1.0\,nm respectively. 

The high-S/N input template spectra are already shifted into the rest frame with wavelength given in air, with the velocity shift determined from the fitted peak of a cross-correlation function with the PHOENIX-ACES spectra for the corresponding stellar parameters. Therefore, in all our abundance fits, the radial velocity ($v_{\rm r}$) was set to 0\,km\,s$^{-1}$. 

\begin{figure*}
    \centering
    \includegraphics[width=0.95\textwidth]{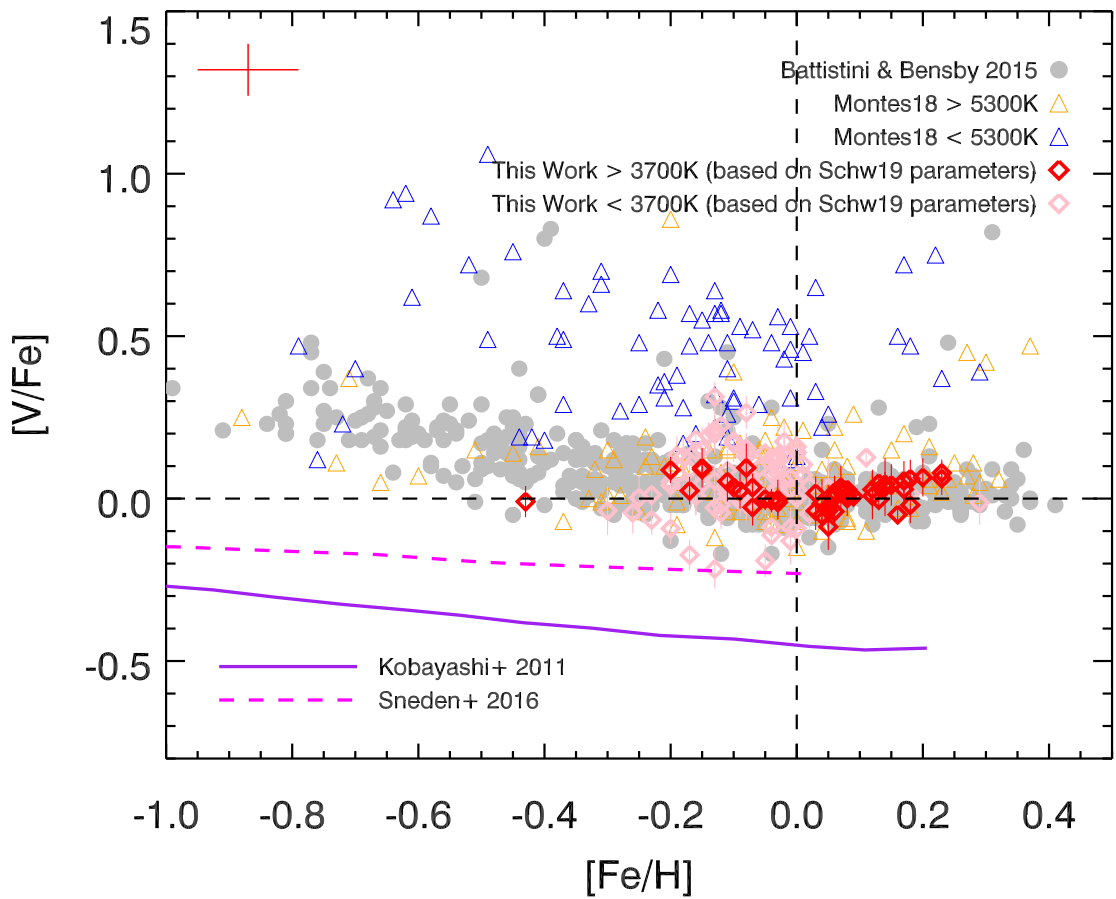}
    \caption{[V/Fe] versus [Fe/H] diagram of the 135 CARMENES GTO stars derived from our line-by-line analysis (red diamonds for $T_{\rm eff}> 3700$\,K and pink diamonds for $T_{\rm eff}< 3700$\,K). Each vertical error bar represents the sample standard deviation of the abundances derived from the set of V lines used for each star. The typical error in the literature metallicity used in our analysis is 0.16\,dex (Schw19), also shown in the upper left of the figure. 
    Overplotted are selected representative V measurements in populations of nearby FGK stars from the literature. Gray filled circles are measurements from \citet{BB15} for local FG dwarfs. The triangles show results from \citet{Montes18} for local FGK dwarfs, with blue denoting the K-dwarf subsample therein ($T_{\rm eff} < 5300\,$K). The V abundances measured for these K dwarfs are generally significantly higher than for the rest of the stars, possibly a consequence of not having accounted for HFS in the modeling. The purple line shows the predicted trend for the solar neighborhood from the Galactic chemical evolution models of \citet{Kobayashi11}. The dashed magenta line shows a model from \citet{Sneden16}, which is built upon the \citet{Kobayashi11} model but incorporates a hypernova jet effect. }
    \label{f:schw19-lbyl_vfe-vs-feh}
\end{figure*}

We present the mean V abundances derived from our individual line-by-line fits in Table~\ref{t:v-abunds}, given in terms of [V/H]\footnote{We used the standard definition $[{\rm X/H}] \equiv \log \epsilon({\rm X}) - \log \epsilon_\odot({\rm X})$, where $\log \epsilon({\rm X}) \equiv \log_{10} (N_{\rm X}/N_{\rm H}) + 12$ and $\log \epsilon_\odot({\rm V}) = 3.93$ \citep{Asplund09}.}. 
Figure~\ref{f:schw19-lbyl_vfe-vs-feh} shows the V-to-Fe abundance ratio of this sample as a function of metallicity (i.e., [Fe/H] from Schw19). Since the [V/Fe] value is not directly constrained in our fits, we calculated it as [V/H] -- [Fe/H]. Our [V/Fe] values lie close to 0.00\,dex (sample standard deviation = 0.09\,dex), consistent with that of FG-type stars in the solar neighborhood \citep[e.g.,][hereafter BB15]{BB15}. The later-type stars in the sample (i.e., $T_{\rm eff}< 3700$\,K) present a larger scatter than the earlier-type stars (i.e., $T_{\rm eff}> 3700$\, K). This could be a consequence of cooler star V abundances being more sensitive to metallicity errors (see Section \ref{ss:error}) and systematically poorer line fits (see Section \ref{ss:line-systematics}).

\subsection{Error quantification}\label{ss:error}

One standard way to estimate the error in abundance measurements is through the line-to-line scatter. Column $\delta$[V/H] in Table~\ref{t:v-abunds} lists the sample standard deviation of the line abundances over all the lines considered for each star. This is typically 0.03\,dex to 0.07\,dex (typically $\sim$0.04\,dex for $T_{\rm eff} > 3700$\,K and $\sim$0.06\,dex for $T_{\rm eff} < 3700$\,K, see also Fig.~\ref{f:lbyl-systematics}). 
The standard error of the mean (SEM) line abundances 
(sample standard deviation over square root of the number of lines) 
is another commonly used measure of abundance error. For all but two stars in our sample, the SEM lies between 0.01\,dex and 0.03\,dex. 

Errors in [V/H] arising from uncertainties in the input stellar parameters were calculated for two representative stars and listed in Table~\ref{t:error-sensitivity}. To derive these values, we varied the $T_{\rm eff}$, $\log g$, and [Fe/H] one-by-one within their formal $1\sigma$ errors as given by Schw19 (51\,K, 0.07\,dex, and 0.16\,dex, respectively). The total expected errors contributed by the incertitude of stellar parameters is calculated from summing each source of error in quadrature (i.e., they are treated independently from one another). Adding the line-to-line standard deviations, 0.06\,dex and 0.08\,dex are representative total uncertainties in our V measurements for typical M0.0\,V and M2.0\,V stars, respectively. 

%\begin{table}[htb]
\begin{table}[]
\begin{center}
\caption{Abundance sensitivity to uncertainties in stellar parameters (from Schw19) for two representative stars in the sample.}
\label{t:error-sensitivity}
\begin{tabular}{lcc}
\hline\hline
\noalign{\smallskip}
Stellar  & J14257+236W & J03213+799 \\
parameters & ($T_{\rm eff}\sim4000$\,K) & ($T_{\rm eff}\sim3500$\,K) \\
\noalign{\smallskip}
\hline
\noalign{\smallskip}
$T_{\rm eff} \pm 51$ K &  $\pm$0.01 dex &  $\pm$0.02 dex\\
   $\log g \pm 0.07$ dex &  $\pm$0.03 dex &  $\pm$0.02 dex\\
   $\rm{[Fe/H]} \pm 0.16$ dex &  $\pm$0.03 dex &  $\pm$0.05 dex\\
\noalign{\smallskip}
\hline
\noalign{\smallskip}
Total & 0.04 dex & 0.06 dex \\
\noalign{\smallskip}
\hline
\end{tabular}
\end{center}
\end{table}

\subsection{Line-to-line abundance systematics}\label{ss:line-systematics}

\begin{figure*}
    \centering
    \includegraphics[width=0.95\textwidth]{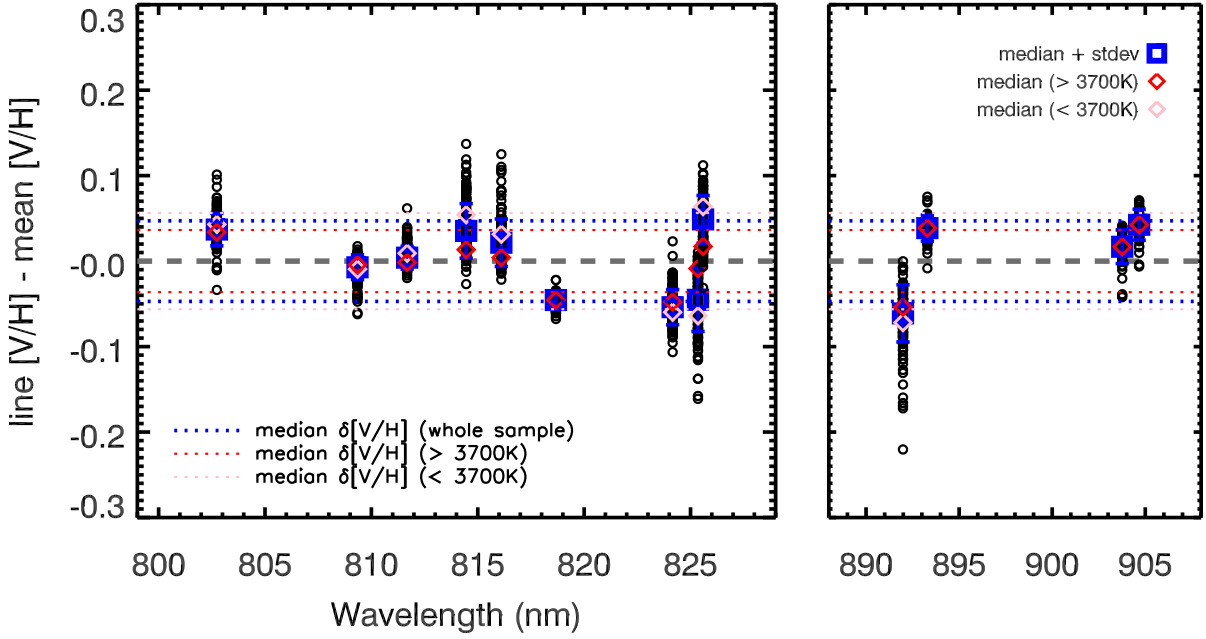}
    \caption{Individual line abundance deviations from the mean abundance across the lines for the star sample. At each given line wavelength (in air), the black circles denote the line [V/H] $-$ mean [V/H] for each star in which that line was studied. The thick orange squares show the median abundance deviation and sample standard deviation for each line (over all the CARMENES stars for which that line was used). The blue and red squares mark the same metric for the warmer ($T_{\rm eff} > 3700$\,K) and cooler ($T_{\rm eff} < 3700$\,K) subsamples, respectively. The horizontal dotted lines bracket the median abundance error (in terms of the line-to-line standard deviation) for the whole sample ($\pm$0.047\,dex, blue squares), the warmer subsample ($\pm$0.036\,dex, red diamonds), and the cooler subsample ($\pm$0.056\,dex, pink diamonds).
    The V~{\sc i} $\lambda$818.6728\,nm, $\lambda$893.2947\,nm, $\lambda$903.7613\,nm, and $\lambda$904.6693\,nm lines were not fitted for stars with $T_{\rm eff} < 3700$\,K.}
    \label{f:lbyl-systematics}
\end{figure*}

We now examine the quality of the line selections and how our final abundance results depend on them. In Fig.~\ref{f:lbyl-systematics} we plot the deviations of individual line abundances from the mean over all lines for the whole sample. On the whole, the abundances derived from our choice of lines show decent agreement. The scatter is typically within $\pm 0.05$ dex, which is comparable to or better than that of similar studies. However, in every line there exists various degrees of systematic deviation, in general worsening for cooler stars. The largest differences are found in the V~{\sc i} $\lambda$818.6728\,nm, $\lambda$824.1599\,nm, and $\lambda$891.9847\,nm lines. Abundances derived from these lines are on average 0.05\,dex above or below the mean across all lines, and can differ more severely for individual stars. 
As shown in Figure~\ref{f:3500k_fitted_lines}, these line regions are more contaminated by molecules and Fe~{\sc i} lines, which could affect the accuracy of the modeling by introducing line blending or complicating the determination of the pseudo-continuum level, as well as make the fits more sensitive to uncertainties in metallicity. The V~{\sc i} $\lambda$814.4560\,nm, $\lambda$816.1062\,nm, $\lambda$825.3506\,nm, and $\lambda$825.5896\,nm lines show good agreement with the average at temperatures higher than $\sim 3700$K, but fare more poorly for cooler stars. This is again indicative of inadequate modeling of molecules. The presence of systematic abundance discrepancies from line to line calls for caution in interpreting abundance constraints from single or small sets of lines unless they are very well-understood. 
  
Examples of single lines that could potentially be used effectively on their own are the V~{\sc i} $\lambda$809.3468\,nm and $\lambda$811.6789\,nm lines. Both appear to trace the mean abundance very well and exhibit very little scatter. The fact that they are useful throughout the $T_{\rm eff}$ range of 3400\,K to 4200\,K makes them the most dependable single V abundance indicator lines in this wavelength and temperature range. By extension, they could be promising metallicity proxies as well (see Section~\ref{ss:metallicity-tracer}).

\subsection{Fits using alternative stellar parameters}\label{ss:alternative-stepars}

Accurate determination of M-dwarf stellar parameters is an active area of research and method-dependent disagreements are commonplace. There is some evidence that the existing published metallicity values for the CARMENES GTO stars (\citealt{Passegger18,Passegger19}, Schw19) derived from PHOENIX model grid fits predominantly to strong Fe, Ti, Mg, and K lines in the optical and NIR, may be somewhat systematically overestimated. For \citet{Passegger19}, who used the latest PHOENIX-SESAM model grids, the systematic offset is especially evident for the earliest M dwarfs and most metal-poor stars (also see a literature comparison in Figs.~5 and~6 of \citealt{Passegger19}). The problem is less severe, but possibly still persists, for the parameters given by Schw19, which were based on PHOENIX-ACES grids. For example, the apparent dearth of relatively metal-poor stars ([Fe/H] $<$ --0.3) in the sample could be suggestive of metallicity measurements being biased to higher values. 
A detailed investigation of such deviations is beyond the scope of this paper.
It may be a combination of saturation effects in the lines, spectral model inadequacies, the neglect of $\alpha$ enhancement, NLTE effects, or something else \citep[see also][for discussions]{Abia20,Ishikawa20,Olander21}. Therefore, it is worthwhile to check the robustness of our V abundances to alternative sets of stellar parameters. 

One further effort to re-determine the fundamental stellar parameters uniformly for the entire CARMENES GTO sample will be presented in a forthcoming publication by \citet{Marfil21}, using the spectral analysis software {\tt{SteParSyn}} \citep{Tabernero18,Tabernero21b}. {\tt{SteParSyn}} is an automatic code designed to infer the stellar atmospheric parameters $T_{\rm eff}$, $\log g$, [Fe/H]\footnote{Denoted as [Fe/H]$_{\rm corr}$ in \citet{Marfil21}.}, and $v\sin i$ of late-type stars (FGKM) following the spectral synthesis method using a Markov Chain Monte Carlo algorithm. This code has already been used in spectroscopic studies that involve late FGKM stars in a variety of astrophysical scenarios \citep[e.g.,][]{Tabernero18, Palle21, Borsa21}. The work of \citet{Marfil21} is based on a synthetic spectral grid generated from BT-Settl model atmospheres \citep{Allard12} coupled to {\tt Turbospectrum} \citep{Plez12} in a large number of Fe~{\sc i} and Ti~{\sc i} line regions, as well as TiO bandheads. 
Currently, some of the $\log g$ values determined with {\tt{SteParSyn}} appear to be possibly too large. Therefore, for our work, we assigned new $\log g$ values based on $T_{\rm eff}$ and [Fe/H] via the PARSEC isochrones \citep{Bressan12,Chen14}, similar to the procedure used by \citet{Passegger19}. The isochrone model inputs were the fitted $T_{\rm eff}$ and [Fe/H] from \citet{Marfil21}, and age estimates as given by \citet{Passegger19}. We present the {\tt{SteParSyn}}-derived $T_{\rm eff}$ and [Fe/H], as well as the $\log g$ constrained from the isochrones, for our CARMENES GTO subsample in Table~\ref{t:v-abunds}. For simplicity, we refer to this collection of alternative stellar parameters, consisting of $T_{\rm eff}$ and [Fe/H] from \citet{Marfil21} and the modified $\log (g)$'s, as the ``Marfil+'' values. 

\begin{figure}
 \centering
    \vspace{-1.3cm}
    \hspace{-0.5cm}
    \includegraphics[width=0.49\textwidth]{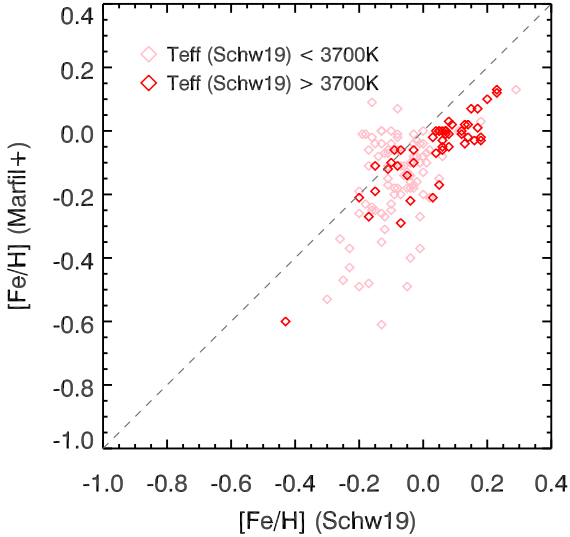}
    \caption{Comparison between [Fe/H] from Schw19 and Marfil+. The pink and red diamonds denote stars with $T_{\rm eff} < 3700$\,K and $T_{\rm eff} > 3700$\,K, respectively. The thin gray dashed line is the one-to-one line.} 
    \label{f:stepar_teff_and_mh}
\end{figure}

As shown in Fig.~\ref{f:stepar_teff_and_mh}, the metallicities given in Marfil+ and in Schw19 systematically differ. 
The discrepancy is largest for relatively metal-poor stars, which also tend to be cooler: those deemed by {\tt{SteParSyn}} to have [Fe/H] $<-0.2$ have on average 0.19\,dex higher value according to Schw19.
Another puzzling pattern is that the majority of stars with super-solar metallicities according to Schw19 are part of the warmer (> 3700\, K) subset, while the metal-poor stars according to Marfil+ are almost exclusively cooler (< 3700\, K). These discrepancies are discussed in more detail by \citet{Marfil21}. 

\begin{figure}
    \centering
    \vspace{-0.5cm}
    \hspace{-0.5cm}
    \includegraphics[width=0.5\textwidth]{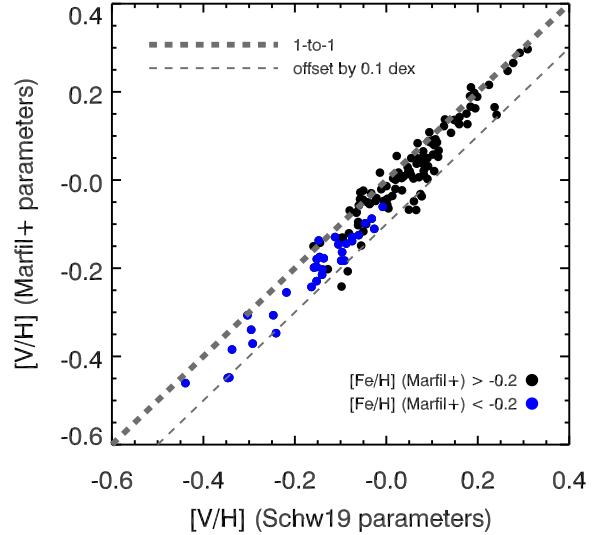}
    \caption{[V/H] measured with stellar parameters fixed to those from Marfil+ versus Schw19. Black and blue points denote stars with [Fe/H] > -0.2 and [Fe/H] < -0.2, respectively. The thick and thin gray dashed lines mark the one-to-one line and that offset by 0.1 dex.} 
    \label{f:vh_schw19_stepar}
\end{figure}

Despite the large systematic 
differences, the final derived [V/H] abundances assuming each set of stellar parameters are very similar. Figure~\ref{f:vh_schw19_stepar} compares the [V/H] values determined from line-by-line fits where $T_{\rm eff}$, $\log g$, and [Fe/H] were fixed to the Marfil+ values with those based on the Schw19 values. Although there is a small systematic offset of 0.034\,dex overall (0.056\,dex for the metal-poor stars with [Fe/H] $< -0.2$), the agreement is within 0.10\,dex in the vast majority of cases, even when the [Fe/H]s differ by up to $\sim$ 0.4 dex.
The relative insensitivity of V abundance derived from the lines used in this work to the exact stellar parameters is not surprising given the error analysis presented in Section~\ref{ss:error}. Therefore, in spite of systematic uncertainties in the exact stellar parameters, our [V/H] values (relative to the solar V abundance of \citealt{Asplund09}) could be taken as robust.       

\begin{figure}
    \centering
    \includegraphics[width=0.5\textwidth]{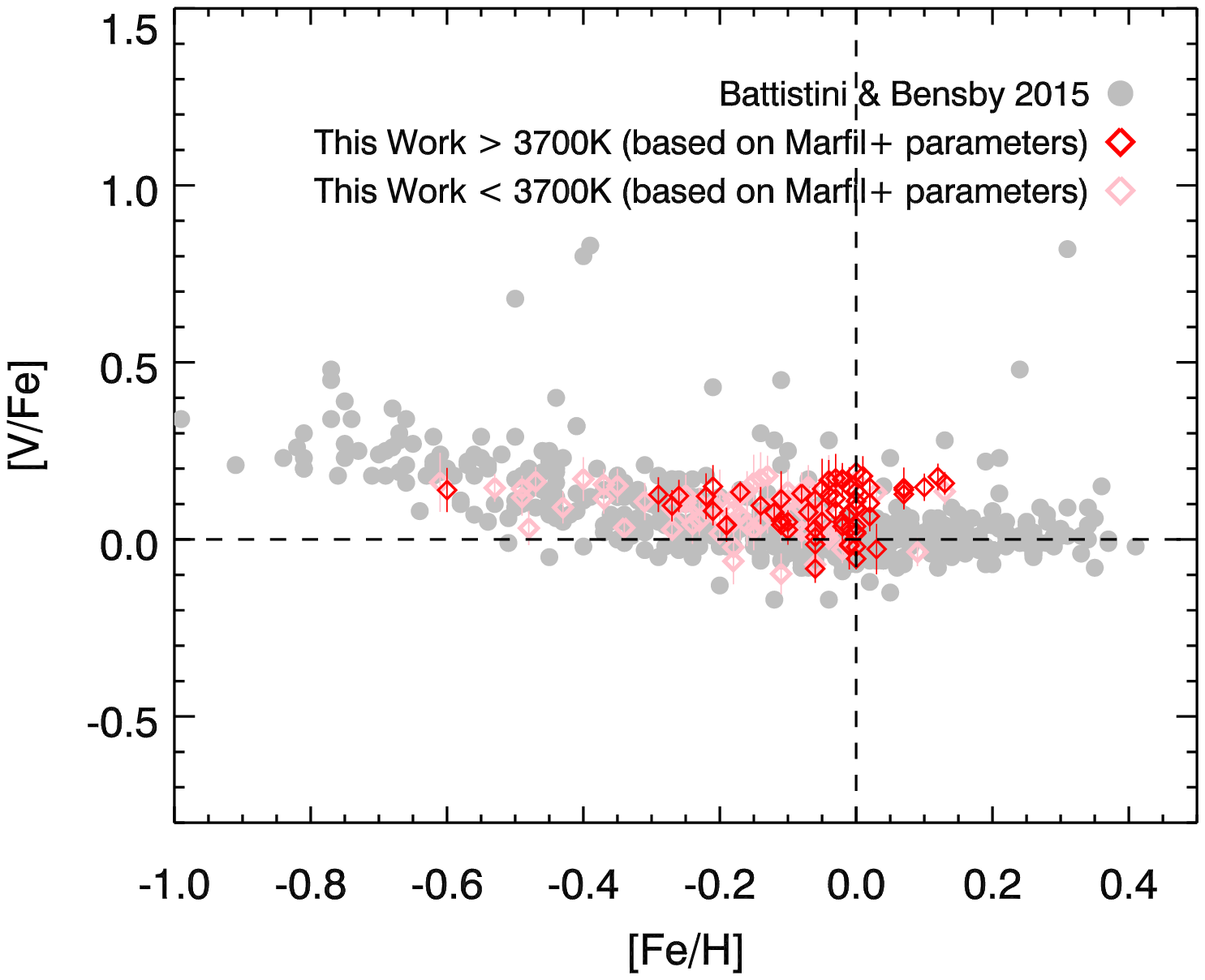}
    \caption{Same as Fig.~\ref{f:schw19-lbyl_vfe-vs-feh}, but for fits based on the Marfil+ stellar parameters for 135 stars. The vertical error bars represent the sample standard deviation of the line-to-line abundances.   
    }
    \label{f:stepar-lbyl_vfe-vs-feh}
\end{figure}

Finally, we show [V/Fe] versus [Fe/H] for the Marfil+ stellar parameters in Figure~\ref{f:stepar-lbyl_vfe-vs-feh}, which exhibits overall good agreement with the distribution of stars from BB15. Since the fitted [V/H] values for the two sets of stellar parameters are very similar, the qualitative differences between Fig.~\ref{f:stepar-lbyl_vfe-vs-feh} and Fig.~\ref{f:schw19-lbyl_vfe-vs-feh} chiefly reflect the differing metallicity measurements between Schw19 and Marfil+. One curious dissimilarity concerns the relative scatter in [V/Fe] between the cooler and warmer subsets: the notably smaller [V/Fe] scatter in the warmer stars in Fig.~\ref{f:schw19-lbyl_vfe-vs-feh} is not reproduced in Fig.~\ref{f:stepar-lbyl_vfe-vs-feh}. These patterns could arise from a complex interplay between the errors from the line fits, systematic uncertainties in the stellar parameters, and perhaps also astrophysical causes. Disentangling these factors is beyond the scope of this work.    

%==================================================================
\section{Discussion}\label{s:discussion}

\subsection{Comparison with local FG dwarfs and K giants}\label{ss:local-fg}

The [V/Fe] versus [Fe/H] trend for the local CARMENES M-dwarf sample is consistent with the behavior observed in populations of local Galactic disk FG-dwarfs stars (e.g., BB15) for the overlapping metallicity range (--0.6\,dex to +0.3\,dex), as shown in Figs.~\ref{f:schw19-lbyl_vfe-vs-feh} and \ref{f:stepar-lbyl_vfe-vs-feh}. 
BB15 performed a comprehensive study of odd-$Z$ iron peak element abundance in the solar neighborhood. In particular, they carried out a large spectroscopic study of 714 FG-dwarfs \citep{Bensby14}, 466 of which have a measured $\log \epsilon({\rm V})$. The analysis was done using a line-by-line approach and differentially with respect to the Sun. 
The authors modeled HFS effects in five V~{\sc i} line regions between 550\,nm and 630\,nm, taking the line component data from \citet{Prochaska00b}. 
An abundance is considered by the authors to be determined if at least one of the five lines appears to be well fit. The BB15 results are also largely compatible with other large studies of nearby stars that have accounted for HFS, including the NIR spectroscopic study of K giants in the SDSS-APOGEE survey of \citet{Hawkins16} and the local giant sample of \citet{Lomaeva19}.

\subsection{A potential partial solution to the issue of abundance-temperature relation in local K dwarfs}\label{ss:k-dwarfs}

A number of well-known abundance studies whose samples also included K dwarfs, many of them dealing with distinguishing the chemical compositions of planet hosts from barren stars, noticed a puzzling systematic trend with regard to V abundance. For example, in a sample of about 1000 FGK stars, \citealt{Adibekyan12} (hereafter AD12) measured [V/Fe] of around 0.0\,dex throughout the FG spectral type range, but rising steadily from 0.0\,dex to +0.6\,dex as $T_{\rm eff}$ decreases from 5300\,K to 4500\,K (as shown in Fig.~\ref{f:kdwarf-sims}). Incidentally, a similar trend is observed therein for Sc, another element known to exhibit HFS (see Section \ref{s:hfs}). Recognizing it to be unphysical, the authors attributed this behavior to a variety of possible causes, including increased line blending for cool stars, NLTE, inaccurate model atmosphere structures, and imprecise $\log(gf)$ values. 

Larger scatter and a systematic correlation with $T_{\rm eff}$ below 5300\,K was also observed in, for example, \citealt{Bodaghee03, Gilli06, Neves09, Tabernero12, Jofre15a}, and \citealt{Montes18}, prompting some authors to either empirically correct the measurements in this temperature range, or to disregard them altogether. This trend is weaker but still visible in
the sample of GK-type giants of \citet{Adibekyan15}. 
In Fig.~\ref{f:schw19-lbyl_vfe-vs-feh}, we overplot the results of \citet{Montes18}, who measured elemental abundances for FGK-type dwarfs in multiple systems with M-dwarf common proper motion companions. 
The distribution of [V/Fe] for stars with $T_{\rm eff} > 5300$\,K and $T_{\rm eff} < 5300$\,K are far removed from each other, with those below 5300\,K enhanced by more than $\sim +0.5$\,dex on average compared to solar. 

\begin{figure*}
    \centering
    \includegraphics[width=\textwidth]{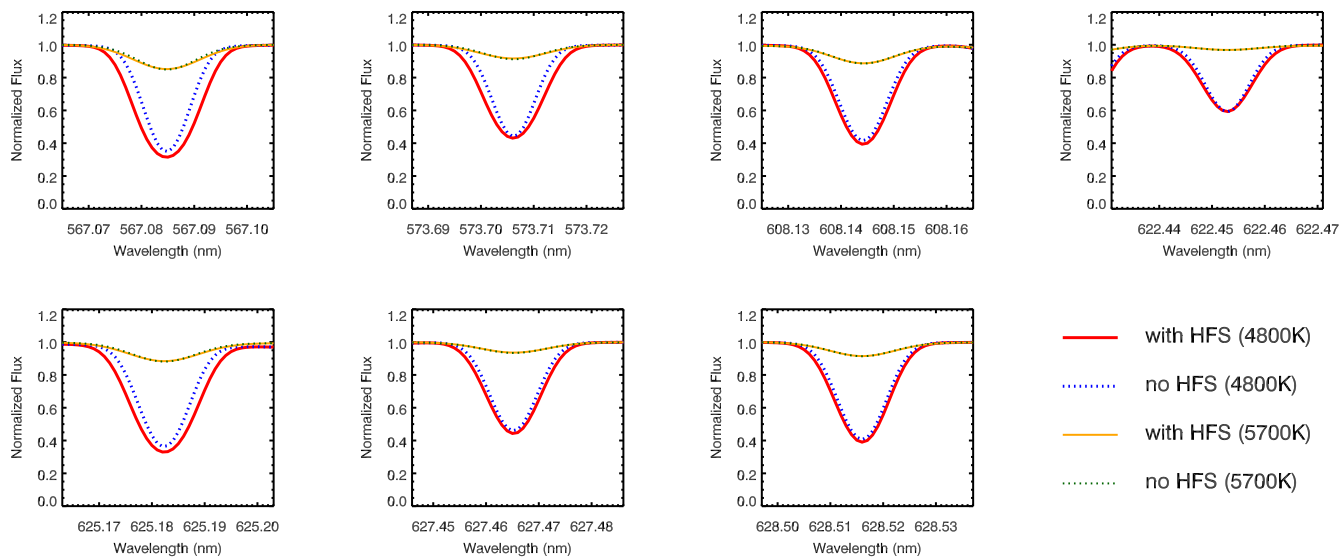}
    \caption{Comparison of simulated profiles of vanadium lines used in \citet{Adibekyan12}, without and with HFS components, for a K dwarf ($T_{\rm eff}$ = 4800\,K) and a G dwarf ($T_{\rm eff}$ = 5700\,K) with a solar abundance pattern at CARMENES VIS resolution ($\mathcal{R} \sim 94\,600$). The red and orange solid lines are generated from the VALD3 atomic line list including the latest HFS components, for the K and G dwarf respectively. The dotted blue and green lines are simulated from a line list without HFS components. The lines are much stronger in later-type stars. While most lines show very small deviations, a few lines (notably $\lambda$567.085\,nm, $\lambda$573.707\,nm, and $\lambda$625.183\,nm) in the K dwarf exhibit significant systematic differences when HFS is considered. }
    \label{f:simulated-spectra-4800k}
\end{figure*}

The oscillator strengths for many V~{\sc i} transitions were improved by \citet{Lawler14}, which could impact works that used older line lists. However, many of the aforementioned analyses were conducted differential with respect to the Sun and, therefore, uncertainties in $\log(gf)$ are expected to cancel. More interesting is the fact that none of the works that found an abundance-temperature trend appear to have used HFS components in their spectral analysis. Inspired by our present work with V lines in cool dwarfs, we hypothesized that a failure to account for HFS could be a key factor responsible for the apparent V overabundance in K dwarfs. Indeed, HFS was already proposed as a possible reason for the large scatter in [V/Fe] of nearby stars compiled from the literature in the Hypatia Catalog \citep{Hinkel14}. A hint could also be found in the work of \citet{Maldonado15}, who also determined elemental abundances in FGK stars. The line regions used in that work for the V analysis were identical to those of \citet{Adibekyan12}, but \citet{Maldonado15} included HFS components for the subset of V~{\sc i} lines redward of 600\,nm \citep{Ramirez14}. The temperature dependence below 5300\,K was still somewhat visible, but far less severe.      

To investigate our hypothesis within our existing analysis framework, we performed a simple experiment. We simulated G- and K-dwarf spectra in the V~{\sc i} line regions used in the abundance work of AD12 with {\tt{iSpec}}. We did this once with the new HFS components, to mimic reality, and once with an older line list likely resembling what would have been used in the previous analysis, to mimic past methodologies. The line profiles are compared in Fig.~\ref{f:simulated-spectra-4800k} for a K dwarf $T_{\rm eff} = 4800$\,K, $\log g = 4.4$, [M/H] = 0.0\,dex, $v\sin i = 1.0$\,km\,s$^{-1}$, and $v_{\rm mic} = 1.0$\,km\,s$^{-1}$, and for a G dwarf with $T_{\rm eff} = 5700$\,K, $\log g = 4.4$, [M/H] = 0.0\,dex, $v\sin i = 1.05$\,km\,s$^{-1}$, and $v_{\rm mic} = 3.92$\,km\,s$^{-1}$. The first thing to note is that, unlike the bucket-shaped lines presented in this work, these line shapes are visually not markedly different from Gaussian and, therefore, they may not raise a red flag. Another thing to note is that all the lines in the K dwarf indeed appear broader (and in some cases also deeper) when HFS is taken into account. Though the effect is subtler in some lines than others, large deviations are apparent for V~{\sc i} $\lambda$567.0848\,nm, $\lambda$573.7062\,nm, and $\lambda$625.1823\,nm. On the other hand, deviations are almost invisible across all the lines in the G dwarf. A third point is that these lines at K dwarf temperatures, $T_{\rm eff} \approx 4800$\,K, are much stronger than in G dwarfs and, therefore, more likely to be affected by saturation, which would mean HFS would no longer conserve their EWs. 

To quantify the impact of the difference between expected and actual line profiles on abundance measurements, we performed line-by-line synthetic spectral fits to the mock spectra generated using HFS components on a temperature grid from 4500\,K to 5500\,K in increments of 100\,K. All mock spectra were synthesized with solar metallicity, $\log g = 4.6$ for $T_{\rm eff} \leq 5000$\,K and $\log g = 4.5$ for $T_{\rm eff} > 5000$\,K, and $v_{\rm mic}$ = 0.88\,km\,s$^{-1}$ to 0.98\,km\,s$^{-1}$, with $v_{\rm mac}$ and $v\sin i$ set to 0. The models used for the abundance fits were generated with a line list ignoring HFS. We ran a trial retrieval using the most current $\log (gf)$ values from \citet{Lawler14}. In addition, we repeated the trials with the $\log (gf)$ data used by \citet{Neves09} (Table~2 therein), which are systematically lower than the updated values by about 10\,\%. In all the fits, only the V abundance was allowed to vary. We took the mean of the abundances retrieved from the seven lines and used the standard error of the mean as an indication of measurement error. Figure~\ref{f:kdwarf-sims} overplots the simulated [V/Fe] versus $T_{\rm eff}$ on the results from AD12. 

The upward trend with decreasing temperature starting from 5300\,K is reproduced by our simulations with a very similar slope and inflection point. Moreover, the worsening line-to-line scatter seen in AD12 also quantitatively resembles that of our simulation. A few ``bad'' lines have a big impact. Exchanging the outdated oscillator strengths with the new values results in a systematic offset of 0.05\,dex to 0.10\,dex, although this difference should only manifest when considering abundance retrievals on an absolute scale. The remaining offset suggests that other sources of systematics still need to be identified, for example NLTE effects or stellar model differences. 
While no dedicated NLTE study of neutral vanadium has been conducted to date, systematic differences between abundances derived from V~{\sc ii} versus V~{\sc i} lines in metal-poor stars do hint that NLTE might be important \citep{Ou20}. It has also been suggested that, at least in Sun-like photospheres, NLTE effects may be significant in V~{\sc i} due to the majority of V atoms being in the singly-ionized state \citep{Scott15b}. The shape and magnitude of the deviation pattern may also differ when using an EW method, as was done by AD12 and similar works. Nevertheless, the results from our simulated SSF abundance retrievals indicate that failure to take HFS into account could play a key role in introducing artificial trends and enhancing the measured V abundances of K dwarfs.   

\begin{figure}
    \centering
    \hspace{-1cm}
    \includegraphics[width=0.5\textwidth]{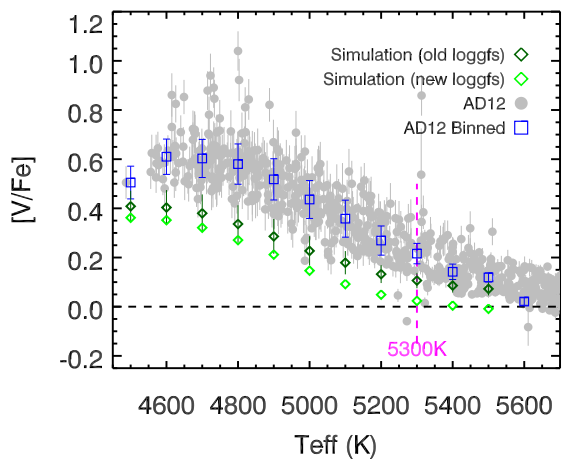}
    \caption{Vanadium abundances based on line-by-line synthetic spectral fitting analysis without HFS on mock spectra generated with an HFS line list as a function of temperature, compared to the abundance trend in AD12. The gray filled circles are measurements and error (computed as line-to-line RMS/$\sqrt{7}$ for seven lines) by AD12 and the blue squares show the median of these measurements and error binned by temperature. The dark green diamonds show the mock retrieval results for model spectra with $\log (gf)$ values from \citet{Neves09} for the V lines, whereas the neon green diamonds use the latest $\log gf$ values. Below 5300\,K, the ``measured'' [V/Fe] increases with decreasing effective temperature in a very similar manner as found by \citet{Adibekyan12} and others. The observed line-to-line scatter as a function of $T_{\rm eff}$ also resembles the simulated behavior.  
    }
    \label{f:kdwarf-sims}
\end{figure}

The effect of unmodeled HFS on vanadium abundance measurements in Sun-like stars was already identified in the appendix of \citet{Takeda07}. They quantified the expected systematic abundance bias from the EW method for nine V~{\sc i} lines between 560\,nm and 660\,nm, two of which are in common with the lines we examined (see Fig.~15 therein). They also found significant corrections to be required for individual lines such as the one at 567.0848\,nm, especially at large strengths. \citet{Takeda07}'s final conclusion that this correction is ``insignificantly small'' (i.e., $\lesssim 0.05$\,dex when averaged over all nine lines) seemingly contradicts our findings. However, considering that their study was limited to cases where $T_{\rm eff} \gtrsim 5000$\,K, the results of \citet{Takeda07} are actually quite compatible with ours. An extrapolation of the steepening trend of the correction factor with $T_{\rm eff}$ below $T_{\rm eff} = 5000$\,K (in their Fig.~16) would imply much larger offsets at cooler temperatures. Further discrepancies could also arise from differences in the exact choice of lines. Relatedly, for red clump giants (4400\,K $< T_{\rm eff} <$ 5100\,K), \citet{Liu07} demonstrated that HFS introduced up to +0.5\,dex of bias in V abundances derived from the three lines that they considered ($\lambda$567.0854\,nm, $\lambda$572.7057\,nm, $\lambda$621.6358\,nm). The deviation worsens with increasing metallicity, presumably because the magnitude of the correction tends to grow with line strength.

\subsection{Trends within the thin and thick disks}\label{ss:thin-thick-disk}

\begin{figure}
\begin{minipage}{0.5\textwidth}
    \centering
    \includegraphics[width=0.9\textwidth]{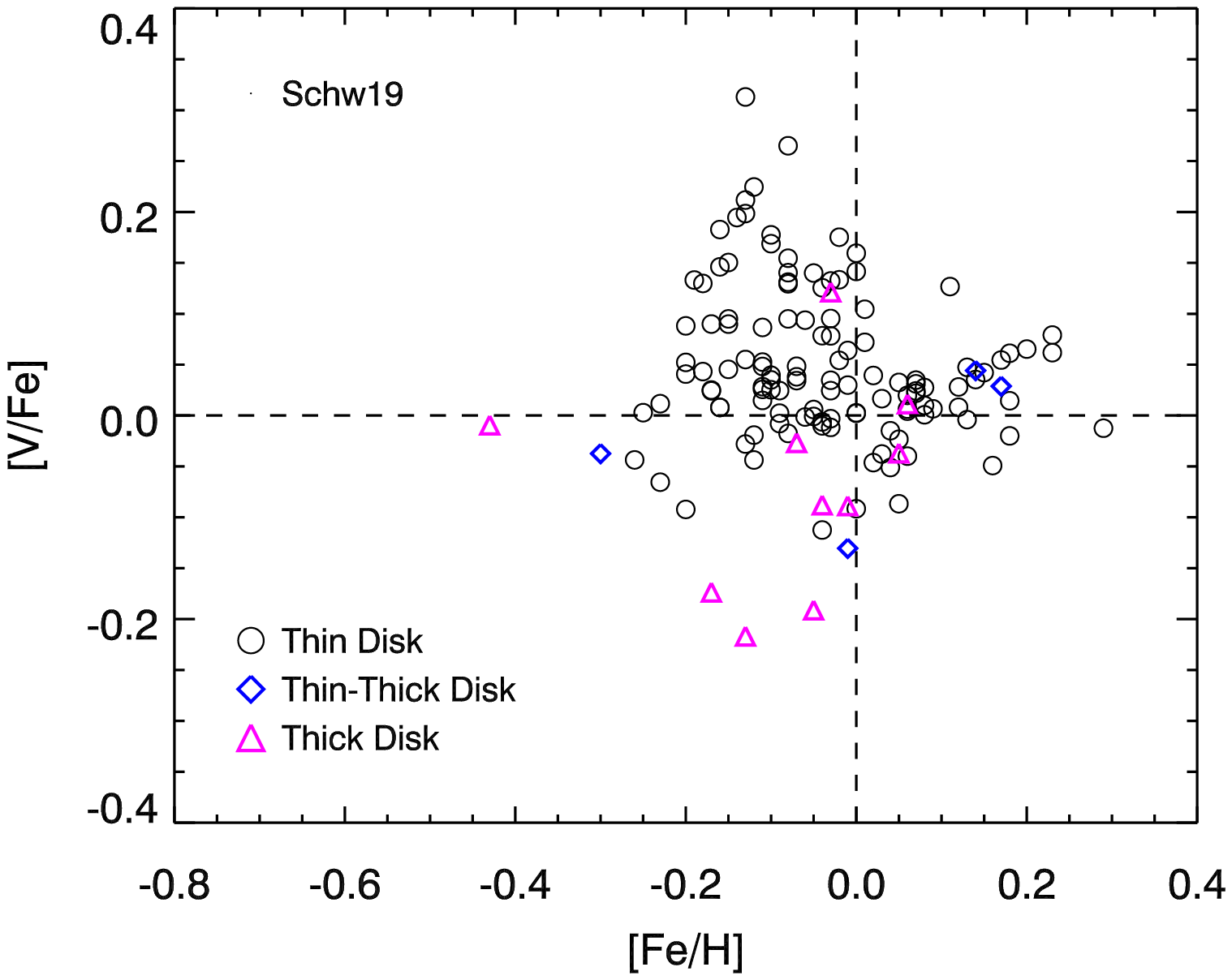}
\end{minipage}
\begin{minipage}{0.5\textwidth}
    \centering
    \includegraphics[width=0.9\textwidth]{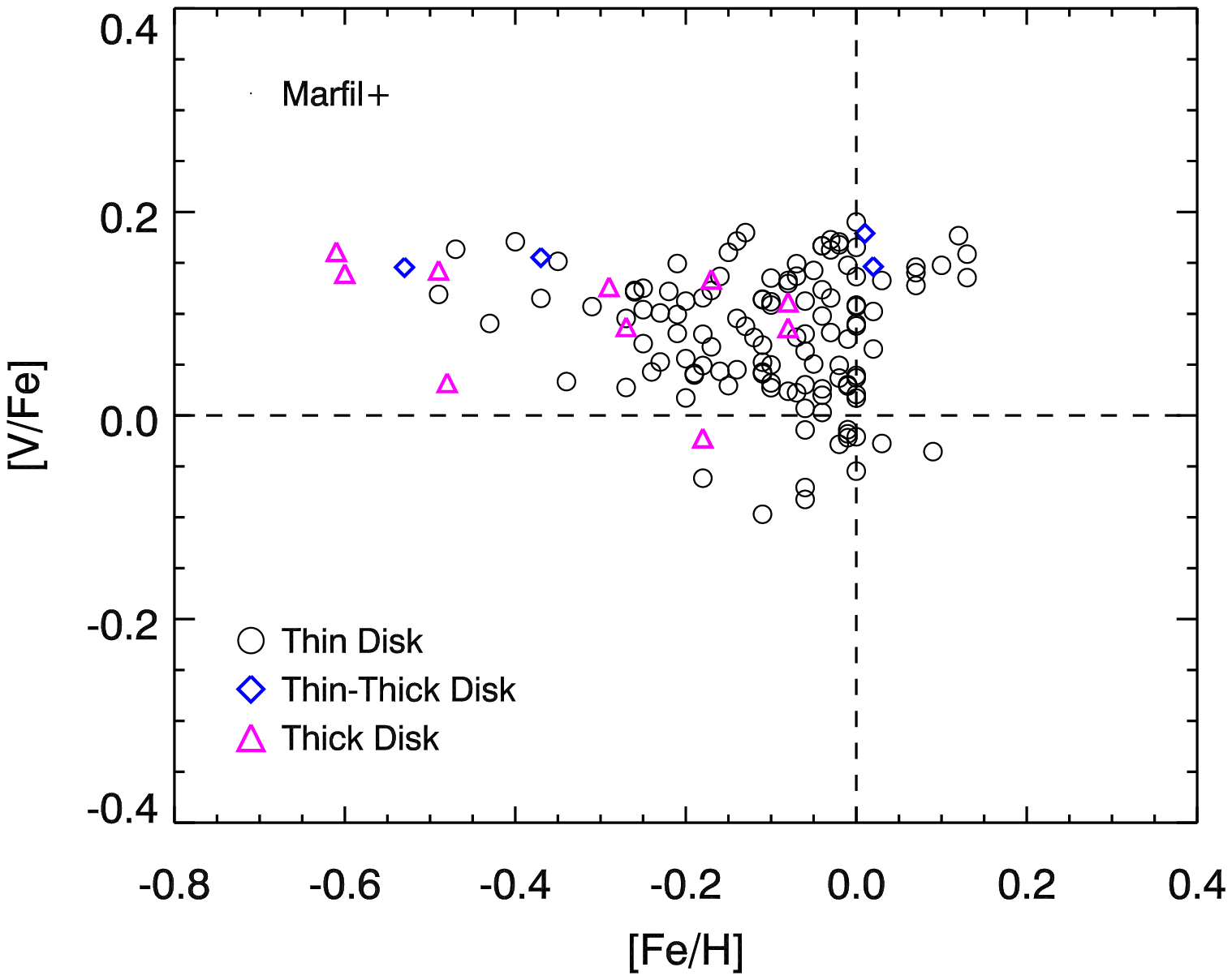}
\end{minipage}
    \caption{[V/Fe] versus [Fe/H] of the 135 CARMENES-GTO stars derived from our line-by-line analysis, color- and symbol-coded for their membership attribution to Galactic kinematic groups (CC21). Top: analysis based on stellar parameters from Schw19. Bottom: analysis based on stellar parameters from Marfil+. The thick-disk members appear to be better distinguished in metallicity space with the parameters estimated by Marfil+. }
    \label{f:schw19-disk_vfe-vs-feh}
\end{figure}

The literature is divided on the existence and nature of fundamental differences in V abundance trends as a function of Galactic population. In a study of ten thick-disk stars, \citet{Prochaska00b} measured an enhancement of about +0.2\,dex in [V/Fe] over solar, which is notably higher than the typical [V/Fe] measurements for thin disk stars available at the time. However, using the same atomic data, \citet{Brewer06} found no significant distinction between V abundance patterns of 23 G dwarfs in the thin and thick disk. Combining data from \citet{Reddy03}, \citet{Reddy06} leveraged $\sim 360$ nearby thin- and thick-disk FG stars to find a $\sim 0.1$\,dex offset between the distribution of [V/Fe] for the thick disk over the thin disk. With 466 FG dwarfs, BB15 showed that [V/Fe] of disk stars in their sample is mostly flat around solar with a slight uptick at the lower end of its metallicity range, whereas a more pronounced sloped relationship of decreasing metallicity and increasing [V/Fe] emerges for the thick disk. While the two segments are visually dissimilar, the authors argued that the thick disk is merely a continuous extension of the thin disk trend into lower metallicities. Recently, \citet{Lomaeva19} resolved a clear, $\sim 0.1$\,dex parallel separation between trends in [V/Fe] versus Fe abundance for $\sim 300$ giant stars belonging to the thin and thick disk (see Fig.~3 therein), even for overlapping metallicity ranges between $-0.4$\,dex and $-0.1$\,dex. We note that the criteria for disk membership assignment are not uniform across the studies. For example, in BB15, the classification was made using isochronal age (with the thin-thick disk boundary drawn at 7--9\,Gyr), while \citet{Lomaeva19} assigned stars to the thin or thick disk based on its [Fe/H], [Ti/Fe], and Galactic kinematics.   

Using kinematic designations from the Carmencita catalog
\citep{Caballero16b,CortesContreras17}, we investigate whether there are thin/thick disk distinctions in V abundance trends in this sample, as shown in Fig.~\ref{f:schw19-disk_vfe-vs-feh}. We mark three different kinematic categories: the thin disk, the thick disk, and the thin-thick disk (i.e., stars with intermediate characteristics, as defined by \citealt{Montes01}). While the Marfil+ [Fe/H] values appear to be better correlated with the kinematic designations than the Schw19 metallicities, we are unable to identify a meaningful trend using either sets of stellar parameters. This could partially be due to small number statistics. All but 14 out of the 135 stars studied here belong to the thin disk.

\subsection{Implication for Galactic nucleosynthesis }\label{ss:nucleosynthesis}

Our measurements reinforce the disagreement between observational and theoretical expectations from GCE models, which predict a significantly lower vanadium-to-iron abundance ratio across this metallicity range. Overplotted on Fig.~\ref{f:schw19-lbyl_vfe-vs-feh} are the relations between [V/Fe] and [Fe/H] in the solar neighborhood from the GCE model of \citet{Kobayashi11} and its variant presented in \citet{Sneden16}. The \citet{Kobayashi11} model is calculated from the then-updated theoretical nucleosynthetic yields for SNe types Ia and II (including hypernovae), as well as for asymptotic giant branch (AGB) stars, in conjunction with observationally constrained initial mass functions (IMFs) and star formation history (SFH) models. The predicted vanadium abundance is around $-0.5$\,dex below that observed for a large range of stellar types and metallicities. Similar deficiencies are also present for Sc and Ti \citep{Kobayashi11,Nomoto13}. In addition to the need to better understand SNe explosion parameters and yields, proposed physics solutions include possible production enhancement facilitated by neutrino interactions \citep{Kobayashi11}, the use of multidimensional nucleosynthetic models \citep{Nomoto13}, incorporation of hypernovae jet effects \citep{Sneden16}, and revised parameters for the IMF and assumptions for SFH and Galaxy formation scenarios \citep{Sneden16}. The \citet{Sneden16} model shown in Fig.~\ref{f:schw19-lbyl_vfe-vs-feh} is based on the \citet{Kobayashi11} model and includes bipolar jets from hypernovae in addition, which reduces the tension with observation to $< 0.3$\,dex. Despite hints of progress, no model to date has satisfactorily reproduced the observed trend. Our work further highlights persistent problems with the current understanding of nucleosynthesis mechanisms and supernova yields for vanadium in the context of the Milky Way's formation history.

\subsection{Potential as a metallicity indicator in cool stars}\label{ss:metallicity-tracer}

\begin{figure}
    \centering
    \hspace{-1cm}
    \includegraphics[width=0.53\textwidth]{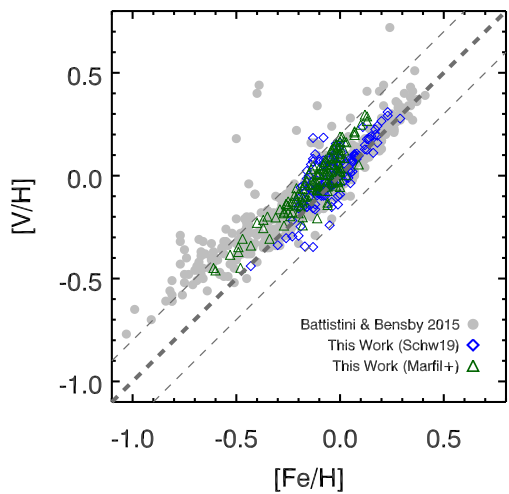}
    \caption{[V/H] versus [Fe/H] measured for stars in the solar neighborhood. The thick gray dashed line is the one-to-one line, while the thin gray dashed lines bracket the $\pm 0.2$\,dex regions. In this metallicity range, [V/H] appears to correlate tightly with [Fe/H], reflecting its potential as an alternative metallicity indicator. }
    \label{f:vh_vs_feh}
\end{figure}

Traditionally, M-dwarf spectra have the notoriety of being riddled with multifarious mysteries (e.g., molecules, magnetic fields, mismatched features). As models grow more sophisticated and data become more exquisite, the question of spectroscopic metallicities for M dwarfs has become the subject of many investigations over the past decade. While many avenues (i.e., spectral ranges and stellar models) have been explored, and progress has been made, modern high-resolution studies still struggle to agree on metallicities for individual cool stars \citep[e.g.,][]{Veyette17,Passegger18, Passegger19, Souto20, Sarmento21} and are also in tension with low-resolution counterparts \citep[e.g.,][]{RA12,Mann13a,Newton14}. Among other implications, the question of metallicity has a significant impact for exoplanet demographics studies \citep[e.g.,][]{Fischer05,Johnson10,Buchhave14,Petigura18}. 

Above [Fe/H] $\sim -0.5$\,dex, V appears to trace Fe abundance in lockstep in a large variety of settings \citep[e.g.,][]{Feltzing98, BB15, Ernandes18}, with dispersion on the order of 0.1\,dex. Figure \ref{f:vh_vs_feh} shows that [V/H] versus [Fe/H] for ours and the BB15 sample of stars in the solar neighborhood fall largely within 0.2\,dex to the one-to-one line. Below [Fe/H] $\sim -0.5$\,dex, [V/H] deviates from an exact correspondence, but maintains a linear correlation with [Fe/H]. In both regimes, the scatter is comparable to or better than precisions currently achievable for metallicity measurements for M-dwarfs as well as study-to-study variations for individual stars. Furthermore, we have demonstrated in Section \ref{s:v-abunds} that [V/H] can be easily and robustly determined nearly independently of other fundamental stellar parameters. Even for cases where a large discrepancy exists between the [Fe/H]s inferred by different procedures (up to $\sim 0.4$ dex), the measured [V/H]s agree within 0.1 dex. Therefore, methods described in this work could represent an alternative, indirect way to estimate the metallicity for M dwarfs. A growing number of high-resolution spectrographs (e.g., 2.2\,m La Silla-FEROS, VLT-UVES, Keck-HIRES, Lick-Hamilton, Gemini-MAROON-X, NEID, HET-HPF, LBT-PEPSI) cover the wavelength range considered in this work. Analyses using observations from these instruments should be capable of delivering reliable vanadium measurements and potentially metallicity proxies for M dwarfs.

\subsection{Implications for planets}\label{ss:planets}

\subsubsection{Planet hosts}\label{sss:planet-hosts}

\begin{figure}
    \begin{minipage}{0.5\textwidth}
    \centering
    \hspace{-1cm}
    \includegraphics[width=0.98\textwidth]{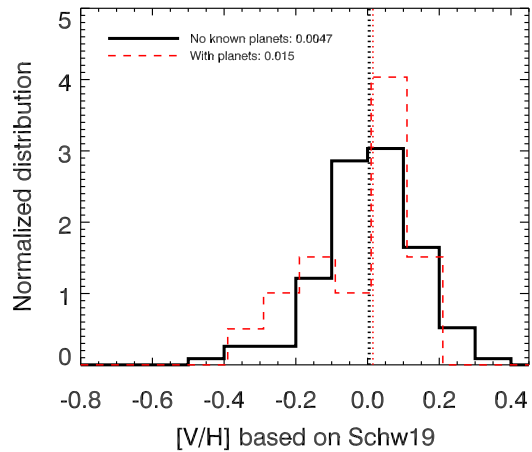}
\end{minipage}
\begin{minipage}{0.5\textwidth}
    \centering
    \hspace{-1cm}
    \includegraphics[width=0.98\textwidth]{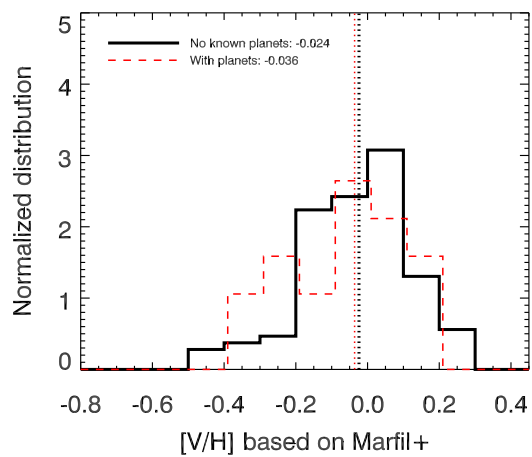}
\end{minipage}
    \caption{Normalized distributions of [V/H] for known planet hosts (red) versus the rest of the stars (black) in the sample based on stellar parameters from Schw19 ({\em top}) and Marfil+ ({\em bottom}). The medians of each distribution are marked by the  vertical dotted lines and given in the legends.}
\label{f:vh_planethosts}
\end{figure}

To date, 19 of the stars in our sample are known planet hosts. As shown in Fig.~\ref{f:vh_planethosts}, the [V/H] distribution of known planet hosts and non-planet hosts do not appear to be significantly different. This is not surprising, as most of the planets hosted by our stars are Neptune-sized or smaller, for which a notable correlation between occurrence rate and the detailed chemical composition of the host star is not expected \citep[e.g.,][]{Adibekyan12}.  

\subsubsection{Planetary atmospheres}\label{sss:planet-atmos}

As better and more precise spectrographs are being developed at large aperture telescopes, V and other atomic species can be searched for in planetary atmospheres. In fact, V has already been detected in at least one planetary atmosphere \citep[WASP-121b][]{Borsa21, Ben-Yami2020}. HFS does not seem to pose a problem for the detection of individual lines in transmission spectroscopy, as the analyses are centered on the stronger lines, and the lines are typically broadened in the planetary atmospheres. However, the use of an HFS line list in the calculation of atomic cross-sections could affect the S/N of the signals. Future studies dealing with retrieving atmospheric abundances of exoplanets might have to take this effect into account. 

It is also interesting to consider how the effect of HFS might manifest in molecules such as VO. While VO, TiO, and metallic species such as Fe~{\sc ii} are believed to be responsible for the thermal inversion of hot exoplanet atmospheres \citep{Yan2020}, there is still no clear detection of VO in a planetary atmosphere. Given that the detection of VO relies on the cross-correlation technique \citep{Snellen2004}, where hundreds of VO spectral lines are cross-correlated to a template, perhaps such methods are more sensitive to HFS effects. However, since the effect of HFS in molecules is more difficult to characterize, and does not appear to be well-studied for VO, a detailed exploration of this topic is beyond the scope of this paper.

%==================================================================
\section{Conclusion}\label{s:conclusion}

Detailed chemical analysis for cool stars stands at the cusp of exploration. In this work, we have performed one of the first elemental abundance studies for a large sample of M dwarfs. In the high-resolution, high S/N, telluric-corrected CARMENES spectra, we located about a dozen hyperfine-split atomic V~{\sc i} lines in the 800--910\,nm region, which have clear presentations and are relatively free of molecular contamination. Using the latest hyperfine components for V~{\sc i} from VALD, we have shown that these lines can be modeled well and that they are sensitive to abundances and relatively insensitive to uncertainties in other fundamental stellar parameters. 
Using spectral synthesis fits to these line regions, we obtained robust [V/H] measurements for 135 nearby, early M dwarfs ($T_{\rm eff} >3400$\,K) in the CARMENES GTO sample, with typical errors of $\sim 0.07$\,dex. 
The [V/Fe] versus [Fe/H] trend closely resembles that observed for local FG-type stars, lying significantly above predictions from GCE models based on the current knowledge of Galactic star formation history and nucleosynthesis channels for this element. Moreover, the tight correlation between Fe- and V- abundances in nearby stars and the relative ease to measure photospheric [V/H] in high-resolution cool star spectra suggests that V has the potential to be an effective metallicity indicator for local M dwarfs.  

We have discussed at length the risks of not properly accounting for HFS, which could bias abundance measurements in various ways. A number of large stellar abundance surveys have reported a systematic trend of increasing [V/Fe] with decreasing $T_{\rm eff}$ for K dwarfs ($T_{\rm eff} \lesssim 5300$\,K). Through simple toy simulations, we have shown that the neglect of HFS is likely responsible for this phenomenon to a large extent. However, the exact impact of HFS on abundance measurements cannot be generalized, as the degree of splitting depends on the exact atomic transition parameters of the lines considered and the properties of the atmosphere. The amount of correction required must be evaluated on a case-by-case basis by considering the stellar type and exact set of lines involved. The safest course of action in any abundance analysis is to use only lines that are well understood (i.e., have complete atomic data available), whenever possible, and to fully model this effect.       

Our investigation underlines the fact that HFS can have a dramatic impact on line profiles and on abundance measurements. We suggest that this quantum effect could play an important role in cool star spectroscopy where a number of elements appear to exhibit stronger and wider splitting patterns than in Sun-like stars. We have exploited this property to provide a reasonable assessment of V abundance in the cool photospheres of the most abundant type of stars in the Galaxy. This type of study is made possible with the accessibility of constantly improving laboratory measurements, which are essential and powerful tools for interpreting increasingly complex stellar spectra in new temperature and wavelength regimes. Together with \citet{Abia20} for Rb, Sr, and Zr, and other forthcoming papers, our findings in the CARMENES spectra open new windows for elemental analysis in cool stars.

%==================================================================
\begin{acknowledgements}
We thank Nikola Vitas and Andrew McWilliam for pedagogical discussions on the nature of HFS splitting. We are grateful to our referee Vardan Adibekyan for insightful feedback that helped to improve the quality of this manuscript. 

  CARMENES is an instrument at the Centro Astron\'omico Hispano-Alem\'an (CAHA) at Calar Alto (Almer\'{\i}a, Spain), operated jointly by the Junta de Andaluc\'ia and the Instituto de Astrof\'isica de Andaluc\'ia (CSIC).
  
  CARMENES was funded by the Max-Planck-Gesellschaft (MPG), 
  the Consejo Superior de Investigaciones Cient\'{\i}ficas (CSIC),
  the Ministerio de Econom\'ia y Competitividad (MINECO) and the European Regional Development Fund (ERDF) through projects FICTS-2011-02, ICTS-2017-07-CAHA-4, and CAHA16-CE-3978, 
  and the members of the CARMENES Consortium 
  (Max-Planck-Institut f\"ur Astronomie,
  Instituto de Astrof\'{\i}sica de Andaluc\'{\i}a,
  Landessternwarte K\"onigstuhl,
  Institut de Ci\`encies de l'Espai,
  Institut f\"ur Astrophysik G\"ottingen,
  Universidad Complutense de Madrid,
  Th\"uringer Landessternwarte Tautenburg,
  Instituto de Astrof\'{\i}sica de Canarias,
  Hamburger Sternwarte,
  Centro de Astrobiolog\'{\i}a and
  Centro Astron\'omico Hispano-Alem\'an), 
  with additional contributions by the MINECO, 
  the Deutsche Forschungsgemeinschaft (DFG) through the Major Research Instrumentation Programme and Research Unit FOR2544 ``Blue Planets around Red Stars'', 
  the Klaus Tschira Stiftung, 
  the states of Baden-W\"urttemberg and Niedersachsen, 
  and by the Junta de Andaluc\'{\i}a.
  
We acknowledge financial support from the Agencia Estatal de Investigaci\'on of the Ministerio de Ciencia, Innovaci\'on y Universidades and the ERDF through projects 
  PID2019-109522GB-C5[1:4]/AEI/10.13039/501100011033     
and the Centre of Excellence ``Severo Ochoa'' and ``Mar\'ia de Maeztu'' awards to the Instituto de Astrof\'isica de Canarias (CEX2019-000920-S), Instituto de Astrof\'isica de Andaluc\'ia (SEV-2017-0709), and Centro de Astrobiolog\'ia (MDM-2017-0737), the Generalitat de Catalunya/CERCA programme,
the Austrian Science Fund (P 33140-N), 
the Ministerio de Universidades (FPU15/01476),
the DFG (SPP 1992 JE 701/5-1),
and
NASA (NNX17AG24G).

This work has made use of the VALD database, operated at Uppsala University, the Institute of Astronomy RAS in Moscow, and the University of Vienna.
\end{acknowledgements}

\bibliographystyle{aa}
\bibliography{refs}

%\clearpage
\begin{appendix}

%\clearpage
\section{Long Table}

Table \ref{t:v-abunds} is available in its entirety in electronic form at the CDS via anonymous ftp to cdsarc.u-strasbg.fr (130.79.128.5) or via http://cdsweb.u-strasbg.fr/cgi-bin/qcat?J/A+A/. 

%\begin{center}
%\begin{longtable}{lccccrrcccrrcr}

%\begin{center}
%\begin{longtable}{lccccrrcccrrcr}
%\longtab{
\onecolumn
\begin{landscape}
\begin{longtable}{lccccrrcccrrcr}
\caption{Measured vanadium abundances for the sample}
\label{t:v-abunds}
\\
\hline\hline
Karmn & SpT & $v\sin i$\textsuperscript{a} & \multicolumn{5}{c}{Schw19 Stellar Parameters} & \multicolumn{5}{c}{Marfil+ Stellar Parameters} & Kinematic 
\\
& & (km/s) & $T_{\rm eff}$ (K) & $\log g$ & [Fe/H] & [V/H] & $\delta$[V/H]\textsuperscript{b} & $T_{\rm eff}$ (K) & $\log g$ & [Fe/H] & [V/H] & $\delta$[V/H]\textsuperscript{b} & Membership \\
\hline
\endhead
\hline
\endfoot
  J00051+457 &M1.0V & 2.0 & 3675 & 4.84 &-0.16 &-0.01 & 0.03 & 3759 & 4.72 &-0.02 & 0.02 & 0.04 &   YD\\
  J00183+440 &M1.0V & 2.0 & 3615 & 4.91 &-0.25 &-0.25 & 0.07 & 3613 & 4.86 &-0.47 &-0.31 & 0.06 &    D\\
  J00389+306 &M2.5V & 2.0 & 3537 & 4.89 &-0.05 &-0.04 & 0.06 & 3559 & 4.81 &-0.18 &-0.10 & 0.06 &    D\\
  J01013+613 &M2.0V & 2.0 & 3529 & 4.94 &-0.16 &-0.15 & 0.06 & 3562 & 4.81 &-0.24 &-0.20 & 0.06 &    D\\
  J01025+716 &M3.0V & 2.0 & 3488 & 4.91 & 0.02 & 0.06 & 0.07 & 3523 & 4.76 &-0.06 & 0.00 & 0.07 &    D\\
  J01026+623 &M1.5V & 2.0 & 3805 & 4.69 & 0.13 & 0.13 & 0.04 & 3806 & 4.71 & 0.02 & 0.12 & 0.04 &   YD\\
  J01433+043 &M2.0V & 2.0 & 3532 & 4.91 &-0.09 &-0.09 & 0.06 & 3550 & 4.82 &-0.20 &-0.14 & 0.06 &    D\\
  J01518+644 &M2.5V & 2.0 & 3555 & 4.90 &-0.08 & 0.06 & 0.03 & 3648 & 4.55 &-0.02 &-0.05 & 0.04 &   YD\\
  J02015+637 &M3.0V & 2.0 & 3489 & 4.93 &-0.05 & 0.09 & 0.05 & 3563 & 4.81 &-0.13 & 0.05 & 0.04 &    D\\
  J02123+035 &M1.5V & 2.0 & 3663 & 4.81 &-0.05 &-0.24 & 0.07 & 3595 & 4.87 &-0.49 &-0.35 & 0.04 &   TD\\
  J02222+478 &M0.5V & 2.0 & 3935 & 4.67 & 0.07 & 0.10 & 0.03 & 3899 & 4.67 & 0.00 & 0.09 & 0.03 &    D\\
  J02358+202 &M2.0V & 2.0 & 3585 & 4.89 &-0.13 & 0.08 & 0.04 & 3715 & 4.55 & 0.00 & 0.02 & 0.05 &   YD\\
  J02442+255 &M3.0V & 2.0 & 3472 & 4.95 &-0.06 &-0.06 & 0.07 & 3495 & 4.88 &-0.14 &-0.09 & 0.07 &   YD\\
 J02565+554W &M1.0V & 2.0 & 3896 & 4.66 & 0.18 & 0.16 & 0.03 & 3828 & 4.68 &-0.03 & 0.14 & 0.02 &    D\\
  J03181+382 &M1.5V & 2.0 & 3861 & 4.66 & 0.20 & 0.27 & 0.03 & 3845 & 4.68 & 0.10 & 0.25 & 0.03 &    D\\
  J03213+799 &M2.0V & 2.0 & 3549 & 4.93 &-0.18 &-0.14 & 0.06 & 3592 & 4.80 &-0.23 &-0.18 & 0.06 &    D\\
  J03217-066 &M2.0V & 2.0 & 3555 & 4.92 &-0.15 & 0.00 & 0.06 & 3653 & 4.75 &-0.04 &-0.01 & 0.05 &   YD\\
  J03463+262 &M0.0V & 3.3 & 3993 & 4.65 & 0.12 & 0.13 & 0.03 & 3965 & 4.69 &-0.01 & 0.14 & 0.02 &   YD\\
  J03531+625 &M3.0V & 2.0 & 3488 & 4.93 &-0.04 &-0.13 & 0.07 & 3496 & 4.84 &-0.18 &-0.20 & 0.07 &   TD\\
  J04225+105 &M3.5V & 2.0 & 3445 & 4.95 & 0.00 & 0.16 & 0.05 & 3526 & 4.79 &-0.04 & 0.13 & 0.05 &   YD\\
  J04290+219 &M0.5V & 3.9 & 4199 & 4.59 & 0.23 & 0.31 & 0.02 & 4143 & 4.63 & 0.12 & 0.30 & 0.02 &    D\\
  J04376+528 &M0.0V & 3.4 & 4034 & 4.68 &-0.09 &-0.07 & 0.03 & 4031 & 4.66 &-0.06 &-0.07 & 0.03 &    D\\
  J04376-110 &M1.5V & 2.0 & 3621 & 4.85 &-0.07 &-0.02 & 0.04 & 3651 & 4.81 &-0.11 &-0.04 & 0.04 &    D\\
  J04429+189 &M2.0V & 2.0 & 3574 & 4.89 &-0.10 & 0.07 & 0.04 & 3674 & 4.71 & 0.00 & 0.04 & 0.04 &    D\\
  J04429+214 &M3.5V & 2.0 & 3437 & 4.97 &-0.03 & 0.05 & 0.06 & 3511 & 4.77 &-0.01 & 0.02 & 0.06 &    D\\
  J04538-177 &M2.0V & 2.0 & 3572 & 4.90 &-0.12 &-0.14 & 0.06 & 3573 & 4.88 &-0.31 &-0.20 & 0.06 &    D\\
  J04588+498 &M0.0V & 2.0 & 4022 & 4.65 & 0.09 & 0.10 & 0.03 & 3994 & 4.67 & 0.02 & 0.09 & 0.03 &    D\\
  J05127+196 &M2.0V & 2.0 & 3582 & 4.90 &-0.13 &-0.07 & 0.05 & 3614 & 4.79 &-0.26 &-0.14 & 0.04 &    D\\
  J05314-036 &M1.5V & 2.0 & 3891 & 4.64 & 0.23 & 0.29 & 0.04 & 3871 & 4.65 & 0.13 & 0.29 & 0.03 &    D\\
  J05365+113 &M0.0V & 3.8 & 4074 & 4.65 & 0.05 & 0.03 & 0.03 & 4064 & 4.68 & 0.00 & 0.02 & 0.02 &   YD\\
  J05415+534 &M1.0V & 2.0 & 3875 & 4.69 & 0.08 & 0.09 & 0.04 & 3859 & 4.54 & 0.03 & 0.00 & 0.04 &   YD\\
  J06103+821 &M2.0V & 2.0 & 3521 & 4.92 &-0.11 &-0.06 & 0.06 & 3546 & 4.82 &-0.17 &-0.10 & 0.06 &    D\\
  J06105-218 &M0.5V & 2.0 & 3837 & 4.70 & 0.07 & 0.09 & 0.04 & 3829 & 4.71 & 0.00 & 0.09 & 0.04 &   YD\\
  J06421+035 &M3.5V & 2.0 & 3454 & 4.95 & 0.00 & 0.00 & 0.06 & 3498 & 4.77 &-0.08 &-0.06 & 0.06 &    D\\
  J06548+332 &M3.0V & 2.0 & 3451 & 4.96 &-0.03 &-0.01 & 0.07 & 3528 & 4.86 &-0.17 &-0.05 & 0.06 &   YD\\
  J07044+682 &M3.0V & 2.0 & 3461 & 4.95 &-0.04 & 0.04 & 0.06 & 3543 & 4.82 &-0.10 & 0.01 & 0.06 &    D\\
  J07287-032 &M3.0V & 2.0 & 3466 & 4.94 &-0.03 & 0.00 & 0.06 & 3504 & 4.84 &-0.18 &-0.06 & 0.06 &    D\\
  J07353+548 &M2.0V & 2.0 & 3520 & 4.94 &-0.16 &-0.15 & 0.06 & 3567 & 4.84 &-0.25 &-0.18 & 0.06 &   YD\\
  J07361-031 &M1.0V & 3.1 & 3894 & 4.69 & 0.05 &-0.04 & 0.05 & 3834 & 4.70 & 0.00 &-0.05 & 0.04 &    D\\
  J07393+021 &M0.0V & 2.0 & 4014 & 4.66 & 0.07 & 0.10 & 0.03 & 3991 & 4.53 &-0.01 & 0.02 & 0.03 &   YD\\
  J08161+013 &M2.0V & 2.0 & 3568 & 4.89 &-0.10 &-0.06 & 0.05 & 3586 & 4.80 &-0.25 &-0.12 & 0.05 &    D\\
  J08293+039 &M2.5V & 2.0 & 3585 & 4.88 &-0.08 & 0.07 & 0.04 & 3675 & 4.55 &-0.01 &-0.03 & 0.04 &   YD\\
  J08358+680 &M2.5V & 2.0 & 3461 & 4.98 &-0.11 &-0.10 & 0.07 & 3498 & 4.77 &-0.06 &-0.13 & 0.08 &    D\\
  J09133+688 &M2.5V & 2.0 & 3555 & 4.92 &-0.16 & 0.02 & 0.04 & 3699 & 4.73 & 0.09 & 0.05 & 0.05 &   YD\\
  J09140+196 &M3.0V & 2.0 & 3493 & 4.95 &-0.10 & 0.08 & 0.05 & 3565 & 4.56 &-0.04 &-0.04 & 0.05 &   YD\\
  J09143+526 &M0.0V & 2.0 & 4024 & 4.68 &-0.05 &-0.05 & 0.03 & 3992 & 4.71 &-0.14 &-0.04 & 0.02 &   YD\\
  J09144+526 &M0.0V & 2.3 & 4005 & 4.68 &-0.03 &-0.03 & 0.03 & 3992 & 4.69 &-0.06 &-0.03 & 0.03 &   YD\\
  J09163-186 &M1.5V & 2.0 & 3585 & 4.91 &-0.18 &-0.05 & 0.06 & 3707 & 4.73 &-0.01 &-0.02 & 0.06 &   YD\\
  J09360-216 &M2.5V & 2.0 & 3505 & 4.94 &-0.12 &-0.16 & 0.09 & 3494 & 4.84 &-0.27 &-0.24 & 0.08 &    D\\
  J09411+132 &M1.5V & 2.0 & 3613 & 4.87 &-0.11 &-0.02 & 0.05 & 3695 & 4.73 &-0.07 &-0.05 & 0.04 &   YD\\
  J09425+700 &M2.0V & 2.0 & 3515 & 4.94 &-0.12 & 0.10 & 0.05 & 3574 & 4.77 &-0.07 & 0.08 & 0.04 &   YD\\
  J09468+760 &M1.5V & 2.0 & 3680 & 4.78 &-0.01 &-0.10 & 0.05 & 3628 & 4.78 &-0.27 &-0.18 & 0.05 &   TD\\
  J09511-123 &M0.5V & 2.0 & 3777 & 4.76 &-0.07 &-0.10 & 0.04 & 3737 & 4.74 &-0.29 &-0.16 & 0.04 &   TD\\
  J09561+627 &M0.0V & 2.0 & 3980 & 4.66 & 0.07 & 0.09 & 0.03 & 3957 & 4.69 & 0.00 & 0.11 & 0.02 &   YD\\
  J10023+480 &M1.0V & 2.0 & 3785 & 4.72 & 0.05 & 0.01 & 0.03 & 3758 & 4.73 &-0.17 &-0.04 & 0.02 &   TD\\
  J10122-037 &M1.5V & 2.0 & 3610 & 4.87 &-0.13 & 0.07 & 0.04 & 3722 & 4.73 &-0.04 & 0.08 & 0.04 &   YD\\
  J10167-119 &M3.0V & 2.0 & 3532 & 4.88 & 0.01 & 0.08 & 0.05 & 3561 & 4.81 &-0.10 & 0.04 & 0.05 &    D\\
  J10251-102 &M1.0V & 2.0 & 3780 & 4.72 & 0.06 & 0.08 & 0.05 & 3746 & 4.72 &-0.06 & 0.05 & 0.04 &   YD\\
  J10289+008 &M2.0V & 2.0 & 3579 & 4.88 &-0.09 &-0.10 & 0.05 & 3586 & 4.59 &-0.18 &-0.24 & 0.06 &   YD\\
  J10350-094 &M3.0V & 2.0 & 3469 & 4.94 &-0.02 & 0.03 & 0.06 & 3546 & 4.82 &-0.16 &-0.02 & 0.05 &    D\\
  J10396-069 &M2.5V & 2.0 & 3554 & 4.88 &-0.02 & 0.11 & 0.05 & 3596 & 4.76 &-0.08 & 0.05 & 0.04 &   YD\\
  J11000+228 &M2.5V & 2.0 & 3499 & 4.94 &-0.11 &-0.08 & 0.06 & 3518 & 4.60 &-0.11 &-0.21 & 0.07 &   YD\\
  J11026+219 &M1.0V & 2.6 & 3905 & 4.69 & 0.04 &-0.01 & 0.04 & 3872 & 4.70 & 0.00 &-0.02 & 0.04 &    D\\
  J11054+435 &M1.0V & 2.0 & 3639 & 4.92 &-0.30 &-0.34 & 0.07 & 3639 & 4.92 &-0.53 &-0.38 & 0.06 & TD-D\\
  J11110+304 &M2.0V & 2.0 & 3753 & 4.71 & 0.13 & 0.18 & 0.04 & 3739 & 4.69 &-0.04 & 0.13 & 0.04 &    D\\
  J11126+189 &M1.5V & 2.0 & 3645 & 4.86 &-0.14 & 0.05 & 0.03 & 3735 & 4.70 &-0.08 & 0.05 & 0.04 &    D\\
  J11302+076 &M2.5V & 2.0 & 3513 & 4.91 &-0.03 & 0.07 & 0.05 & 3563 & 4.59 &-0.10 &-0.07 & 0.05 &    D\\
  J11306-080 &M3.5V & 2.0 & 3422 & 4.98 &-0.01 & 0.02 & 0.07 & 3491 & 4.78 &-0.04 &-0.02 & 0.07 &   YD\\
  J11467-140 &M3.0V & 2.0 & 3570 & 4.82 & 0.11 & 0.24 & 0.05 & 3629 & 4.72 & 0.00 & 0.17 & 0.04 &    D\\
  J11511+352 &M1.5V & 2.0 & 3645 & 4.87 &-0.17 &-0.14 & 0.05 & 3704 & 4.70 &-0.06 &-0.14 & 0.06 &    D\\
 J12123+544S &M0.0V & 2.0 & 3933 & 4.70 &-0.03 &-0.04 & 0.03 & 3910 & 4.73 &-0.10 &-0.05 & 0.03 &    D\\
  J12230+640 &M3.0V & 2.0 & 3540 & 4.86 & 0.05 & 0.08 & 0.06 & 3578 & 4.80 &-0.15 & 0.01 & 0.05 &    D\\
  J12248-182 &M2.0V & 2.0 & 3485 & 4.98 &-0.17 &-0.34 & 0.11 & 3476 & 4.93 &-0.48 &-0.45 & 0.08 &   TD\\
  J12312+086 &M0.5V & 2.0 & 3914 & 4.71 &-0.07 &-0.04 & 0.03 & 3919 & 4.67 &-0.06 &-0.05 & 0.03 &    D\\
  J12350+098 &M2.5V & 2.0 & 3572 & 4.85 & 0.02 &-0.03 & 0.07 & 3563 & 4.81 &-0.21 &-0.11 & 0.07 &    D\\
  J12388+116 &M3.0V & 2.0 & 3439 & 4.96 & 0.00 & 0.14 & 0.05 & 3511 & 4.77 & 0.00 & 0.11 & 0.05 &    D\\
  J13196+333 &M1.5V & 2.0 & 3812 & 4.67 & 0.18 & 0.24 & 0.04 & 3792 & 4.54 &-0.02 & 0.15 & 0.03 &   YD\\
  J13209+342 &M1.0V & 2.0 & 3763 & 4.73 & 0.03 &-0.01 & 0.04 & 3760 & 4.73 &-0.21 &-0.06 & 0.04 &    D\\
 J13283-023W &M3.0V & 2.0 & 3482 & 4.92 & 0.01 & 0.11 & 0.07 & 3554 & 4.75 &-0.07 & 0.07 & 0.06 &    D\\
  J13299+102 &M0.5V & 2.0 & 3714 & 4.81 &-0.15 &-0.06 & 0.05 & 3779 & 4.73 &-0.11 &-0.06 & 0.05 &    D\\
  J13450+176 &M0.0V & 2.3 & 3806 & 4.85 &-0.43 &-0.44 & 0.05 & 3816 & 4.83 &-0.60 &-0.46 & 0.04 &   TD\\
  J13457+148 &M1.5V & 2.0 & 3678 & 4.79 &-0.04 &-0.15 & 0.06 & 3619 & 4.86 &-0.40 &-0.23 & 0.04 &    D\\
  J14010-026 &M1.0V & 2.0 & 3724 & 4.77 &-0.04 &-0.05 & 0.04 & 3710 & 4.75 &-0.22 &-0.10 & 0.04 &    D\\
  J14082+805 &M1.0V & 2.0 & 3841 & 4.67 & 0.16 & 0.11 & 0.04 & 3803 & 4.68 &-0.03 & 0.09 & 0.03 &    D\\
  J14251+518 &M2.5V & 2.0 & 3521 & 4.91 &-0.07 &-0.03 & 0.06 & 3558 & 4.81 &-0.20 &-0.09 & 0.06 &    D\\
 J14257+236E &M0.5V & 2.0 & 3947 & 4.65 & 0.17 & 0.20 & 0.03 & 3907 & 4.67 & 0.01 & 0.19 & 0.02 & TD-D\\
 J14257+236W &M0.0V & 2.0 & 4024 & 4.64 & 0.17 & 0.22 & 0.03 & 3983 & 4.67 & 0.07 & 0.22 & 0.02 &    D\\
  J14294+155 &M2.0V & 2.0 & 3671 & 4.76 & 0.06 & 0.07 & 0.04 & 3653 & 4.72 &-0.08 & 0.01 & 0.04 &   TD\\
  J14307-086 &M0.5V & 2.4 & 4089 & 4.63 & 0.14 & 0.18 & 0.02 & 4026 & 4.66 & 0.02 & 0.17 & 0.01 & TD-D\\
  J14524+123 &M2.0V & 2.0 & 3548 & 4.89 &-0.08 & 0.19 & 0.04 & 3714 & 4.70 & 0.07 & 0.21 & 0.05 &    D\\
  J15095+031 &M3.0V & 2.0 & 3492 & 4.92 &-0.01 & 0.05 & 0.06 & 3545 & 4.82 &-0.10 & 0.01 & 0.05 &    D\\
  J15194-077 &M3.0V & 2.0 & 3441 & 4.98 &-0.08 &-0.10 & 0.09 & 3488 & 4.88 &-0.18 &-0.13 & 0.07 &   YD\\
  J15598-082 &M1.0V & 2.0 & 3625 & 4.89 &-0.19 &-0.06 & 0.04 & 3728 & 4.70 &-0.01 &-0.03 & 0.05 &    D\\
 J16167+672N &M3.0V & 2.0 & 3488 & 4.93 &-0.03 & 0.10 & 0.08 & 3575 & 4.74 &-0.04 & 0.06 & 0.07 &   YD\\
 J16167+672S &M0.0V & 2.7 & 4090 & 4.62 & 0.15 & 0.19 & 0.04 & 4054 & 4.66 & 0.07 & 0.20 & 0.03 &    D\\
  J16254+543 &M1.5V & 2.0 & 3525 & 4.97 &-0.26 &-0.30 & 0.08 & 3588 & 4.90 &-0.34 &-0.31 & 0.06 &   YD\\
  J16462+164 &M2.5V & 2.0 & 3506 & 4.92 &-0.06 & 0.03 & 0.05 & 3558 & 4.81 &-0.11 & 0.00 & 0.05 &    D\\
  J16581+257 &M1.0V & 2.0 & 3734 & 4.78 &-0.08 & 0.01 & 0.04 & 3780 & 4.73 &-0.11 & 0.00 & 0.04 &    D\\
  J17052-050 &M1.5V & 2.0 & 3642 & 4.81 &-0.01 &-0.14 & 0.07 & 3597 & 4.87 &-0.37 &-0.21 & 0.06 & TD-D\\
  J17166+080 &M2.0V & 2.0 & 3543 & 4.91 &-0.10 &-0.07 & 0.06 & 3567 & 4.81 &-0.23 &-0.13 & 0.06 &   YD\\
  J17303+055 &M0.0V & 2.0 & 3812 & 4.77 &-0.15 &-0.06 & 0.03 & 3825 & 4.57 &-0.19 &-0.15 & 0.03 &   YD\\
  J17355+616 &M0.5V & 2.0 & 3893 & 4.69 & 0.06 & 0.07 & 0.05 & 3878 & 4.67 & 0.00 & 0.04 & 0.05 &    D\\
  J17378+185 &M1.0V & 2.0 & 3656 & 4.87 &-0.23 &-0.22 & 0.05 & 3675 & 4.83 &-0.37 &-0.25 & 0.05 &    D\\
  J18051-030 &M1.0V & 2.0 & 3663 & 4.86 &-0.20 &-0.16 & 0.06 & 3725 & 4.74 &-0.19 &-0.15 & 0.06 &    D\\
  J18198-019 &K7.0V & 2.0 & 4143 & 4.66 &-0.11 &-0.06 & 0.02 & 4162 & 4.69 &-0.12 &-0.04 & 0.02 &   YD\\
  J18353+457 &M0.5V & 2.0 & 3932 & 4.68 & 0.06 & 0.02 & 0.03 & 3896 & 4.67 &-0.05 & 0.00 & 0.03 &    D\\
  J18409-133 &M1.0V & 2.0 & 3812 & 4.70 & 0.08 & 0.11 & 0.04 & 3782 & 4.69 &-0.05 & 0.09 & 0.03 &    D\\
  J18480-145 &M2.5V & 2.0 & 3496 & 4.94 &-0.11 &-0.08 & 0.07 & 3540 & 4.82 &-0.15 &-0.12 & 0.07 &    D\\
  J18580+059 &M0.5V & 2.0 & 3923 & 4.68 & 0.08 & 0.08 & 0.03 & 3898 & 4.67 &-0.01 & 0.07 & 0.03 &    D\\
  J19070+208 &M2.0V & 2.0 & 3536 & 4.94 &-0.20 &-0.29 & 0.11 & 3541 & 4.90 &-0.49 &-0.37 & 0.08 &    D\\
  J19072+208 &M2.0V & 2.0 & 3526 & 4.96 &-0.23 &-0.30 & 0.10 & 3576 & 4.88 &-0.43 &-0.34 & 0.08 &    D\\
 J19169+051N &M2.5V & 2.0 & 3534 & 4.90 &-0.04 & 0.09 & 0.05 & 3587 & 4.80 &-0.14 & 0.03 & 0.04 &    D\\
  J19346+045 &M0.0V & 3.9 & 4062 & 4.69 &-0.17 &-0.15 & 0.04 & 4047 & 4.68 &-0.27 &-0.17 & 0.03 &    D\\
  J20450+444 &M1.5V & 2.0 & 3586 & 4.89 &-0.15 &-0.10 & 0.07 & 3615 & 4.82 &-0.25 &-0.15 & 0.06 &   YD\\
  J20533+621 &M1.0V & 2.0 & 3843 & 4.71 & 0.04 & 0.03 & 0.04 & 3815 & 4.68 &-0.07 & 0.01 & 0.03 &    D\\
  J20567-104 &M2.5V & 2.0 & 3514 & 4.91 &-0.03 & 0.09 & 0.05 & 3571 & 4.74 &-0.08 & 0.03 & 0.05 &   TD\\
  J21019-063 &M2.5V & 2.0 & 3507 & 4.92 &-0.08 & 0.05 & 0.04 & 3560 & 4.75 &-0.06 & 0.02 & 0.04 &    D\\
  J21152+257 &M3.0V & 2.0 & 3675 & 4.68 & 0.29 & 0.28 & 0.05 & 3700 & 4.67 & 0.13 & 0.27 & 0.04 &    D\\
  J21164+025 &M3.0V & 2.0 & 3485 & 4.94 &-0.08 & 0.05 & 0.06 & 3552 & 4.60 &-0.11 &-0.07 & 0.06 &   YD\\
  J21221+229 &M1.0V & 2.0 & 3705 & 4.83 &-0.20 &-0.11 & 0.05 & 3755 & 4.73 &-0.21 &-0.13 & 0.05 &    D\\
  J21348+515 &M3.0V & 2.0 & 3497 & 4.91 & 0.00 & 0.00 & 0.06 & 3523 & 4.86 &-0.13 &-0.04 & 0.06 &   YD\\
  J22021+014 &M0.5V & 2.0 & 3915 & 4.69 & 0.03 & 0.05 & 0.03 & 3904 & 4.67 &-0.02 & 0.03 & 0.03 &    D\\
  J22057+656 &M1.5V & 2.0 & 3653 & 4.86 &-0.17 &-0.08 & 0.04 & 3727 & 4.74 &-0.11 &-0.07 & 0.05 &    D\\
  J22096-046 &M3.5V & 2.0 & 3467 & 4.94 &-0.02 & 0.16 & 0.05 & 3579 & 4.77 &-0.03 & 0.13 & 0.05 &   YD\\
  J22115+184 &M2.0V & 2.0 & 3550 & 4.93 &-0.13 & 0.18 & 0.03 & 3723 & 4.70 & 0.00 & 0.19 & 0.04 &    D\\
  J22125+085 &M3.0V & 2.0 & 3501 & 4.92 &-0.04 &-0.05 & 0.07 & 3513 & 4.83 &-0.16 &-0.12 & 0.07 &    D\\
  J22330+093 &M1.0V & 2.0 & 3669 & 4.86 &-0.20 &-0.15 & 0.04 & 3732 & 4.78 &-0.26 &-0.14 & 0.04 &   YD\\
  J22503-070 &M0.5V & 2.0 & 3904 & 4.72 &-0.10 &-0.06 & 0.03 & 3911 & 4.70 &-0.10 &-0.07 & 0.03 &    D\\
  J22559+178 &M1.0V & 2.0 & 3834 & 4.70 & 0.06 & 0.06 & 0.04 & 3808 & 4.68 &-0.03 & 0.05 & 0.04 &    D\\
  J22565+165 &M1.5V & 2.0 & 3805 & 4.69 & 0.12 & 0.15 & 0.04 & 3787 & 4.69 & 0.00 & 0.14 & 0.03 &    D\\
  J23245+578 &M1.0V & 2.0 & 3842 & 4.68 & 0.14 & 0.18 & 0.03 & 3803 & 4.68 &-0.02 & 0.15 & 0.02 &    D\\
  J23340+001 &M2.5V & 2.0 & 3559 & 4.86 & 0.00 &-0.09 & 0.06 & 3531 & 4.83 &-0.20 &-0.18 & 0.05 &    D\\
  J23381-162 &M2.0V & 2.0 & 3551 & 4.91 &-0.13 &-0.16 & 0.08 & 3583 & 4.91 &-0.35 &-0.20 & 0.07 &    D\\
  J23492+024 &M1.0V & 2.0 & 3653 & 4.84 &-0.13 &-0.35 & 0.09 & 3572 & 4.96 &-0.61 &-0.45 & 0.06 &   TD\\
  J23556-061 &M2.5V & 2.0 & 3694 & 4.71 & 0.18 & 0.19 & 0.04 & 3698 & 4.70 & 0.03 & 0.16 & 0.03 &    D\\
\hline\hline
\end{longtable}
\tablefoot{ \\
\tablefoottext{a}{\textsuperscript{a}: the $v\sin i$'s given here are the rotational broadening parameters used for each star in the spectral fits. It is set to 2.0 km/s when the measurement (from \citealt{Reiners18}) only gives an upper limit of 2.0 km/s, i.e., the true $v \sin i \leq 2$ km/s.} \\   
\tablefoottext{b}{\textsuperscript{b}: [V/H] error is given as the sample standard deviation of the fitted abundances over all line regions used for that particular star.}
}
\end{landscape}
%}

%\clearpage
\section{Representative CARMENES spectra in the wavelength region 800--910\,nm}

\begin{figure*}
    \centering
    \hspace{-1cm}
    \includegraphics[width=\textwidth]{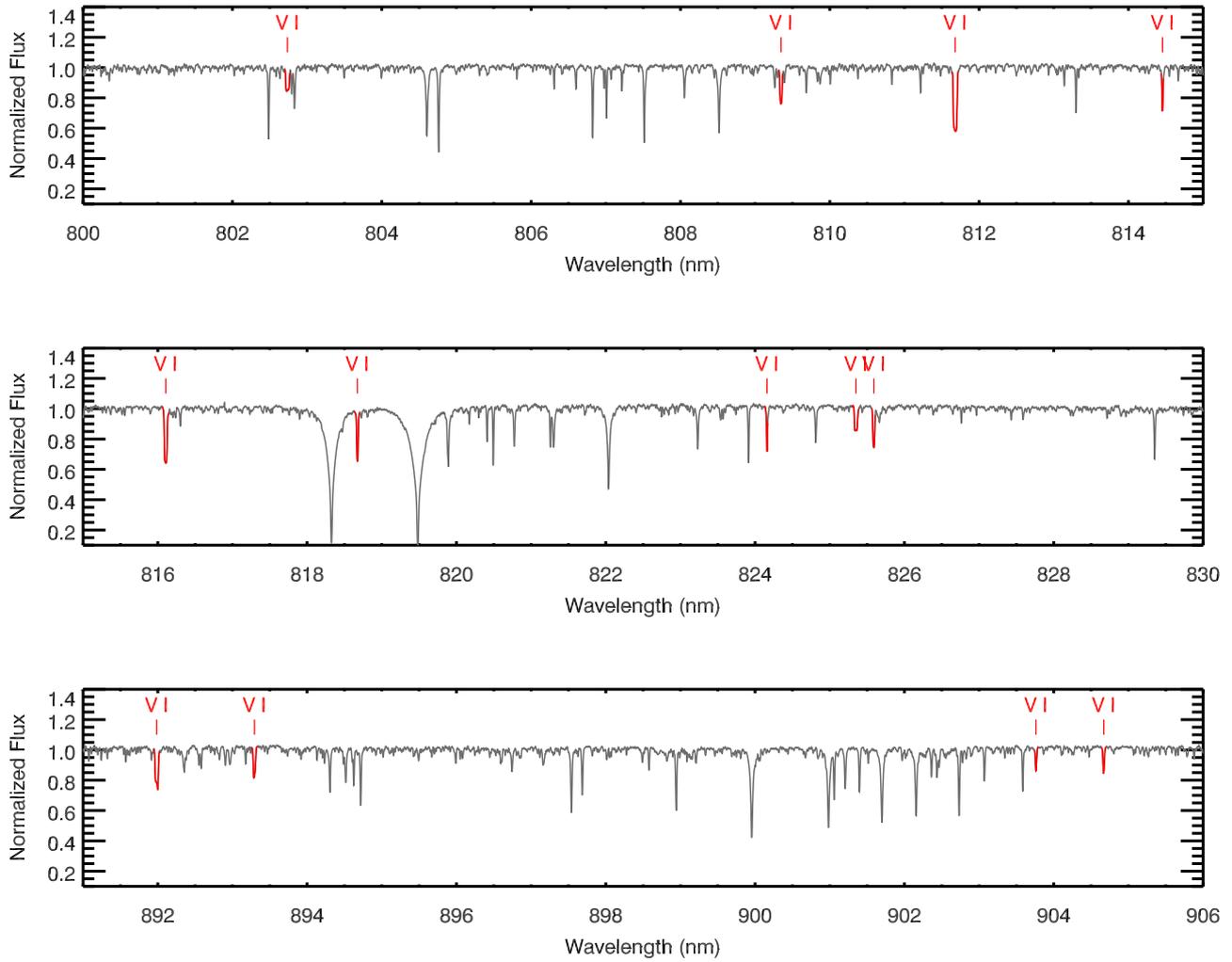}
    \caption{View of a representative, normalized, telluric-corrected and co-added CARMENES spectrum in the wavelength ranges considered in this work for a typical M0.0\,V star (J14257+236W). The V~{\sc i} lines used in this study are marked in red. }
    \label{f:4000k_representative_spectrum}
\end{figure*}

\begin{figure*}
    \centering
    \hspace{-1cm}
    \includegraphics[width=\textwidth]{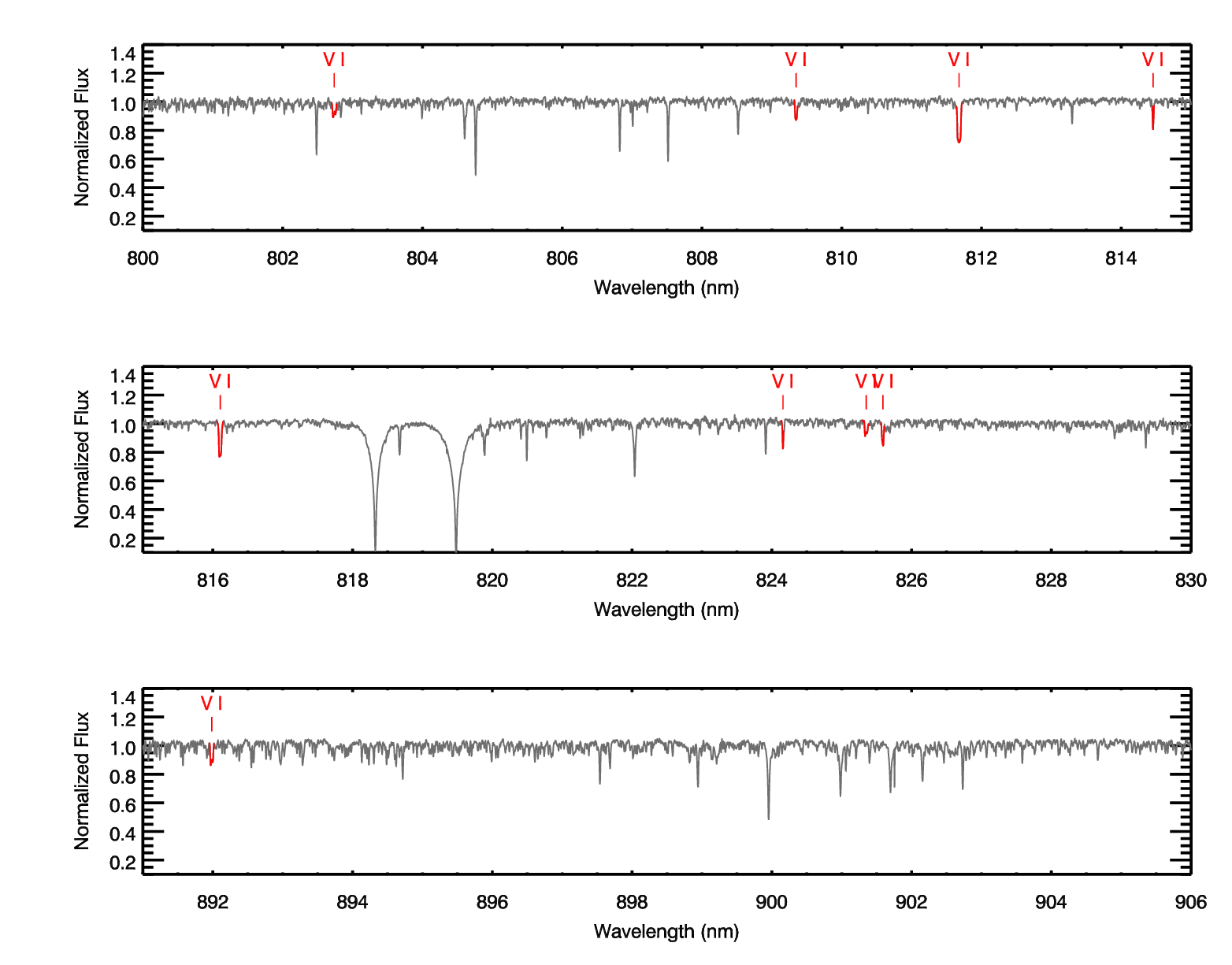}
    \caption{Same as Fig.~\ref{f:4000k_representative_spectrum}, but for a typical M2.0\,V star (J03213+799). }
    \label{f:3500k_representative_spectrum}
\end{figure*}

%\clearpage
\section{All fitted vanadium line regions}\label{a:lineregs}

\begin{figure*}
    \centering
    \hspace{-1cm}
    \includegraphics[width=\textwidth]{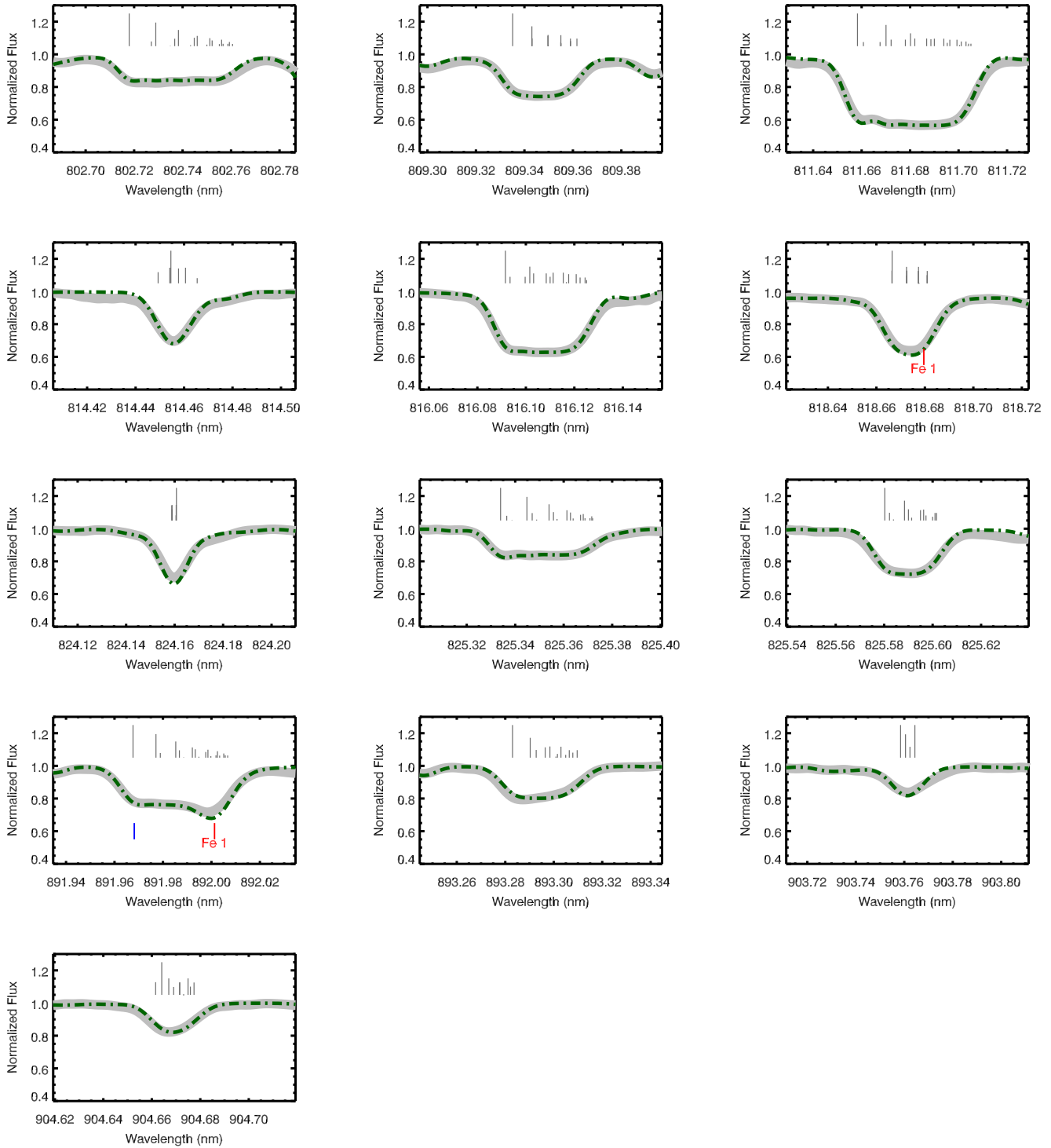}
    \caption{All V~{\sc i} line regions fitted to the observed CARMENES spectrum (thick gray) of a typical M0.0\,V star (J14257+236W). The model corresponding to the best-fit V abundance, using the Schw19 stellar parameters, is shown in green dot-dashed lines. Individual V~{\sc i} HFS components and their relative strengths, as given by $\log(gf\lambda) - \theta\chi$, are shown in vertical gray lines above each feature. Notable contaminating lines (i.e., expected depth $> 0.05$) inside the V~{\sc i} features from other species are marked in blue (for TiO) and red. }
    \label{f:4000k_fitted_lines}
\end{figure*}

\begin{figure*}
    \centering
    \hspace{-1cm}
    \includegraphics[width=\textwidth]{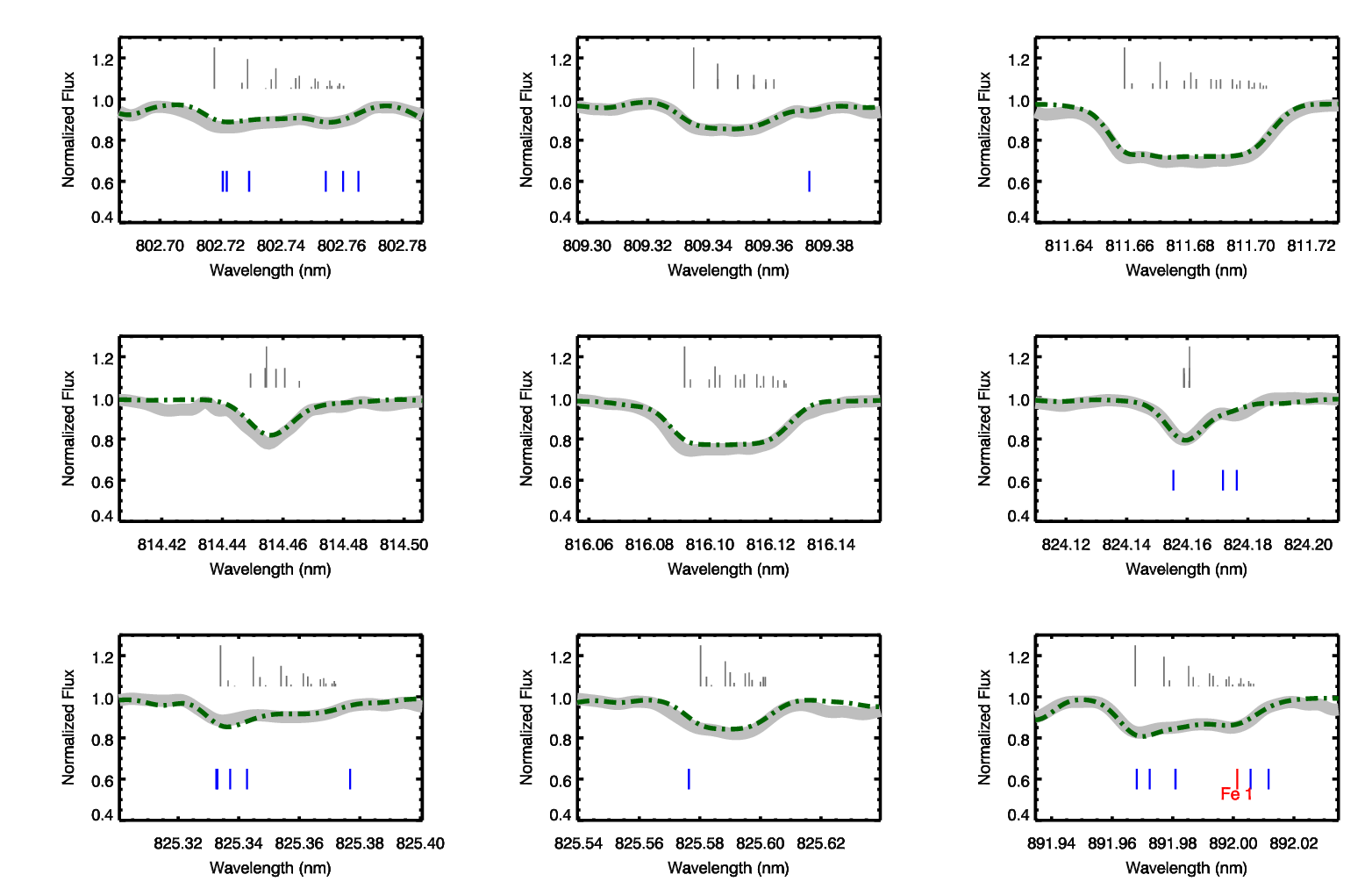}
    \caption{Same as Fig.~\ref{f:4000k_fitted_lines}, but for a typical M2.0\,V star (J03213+799). }
    \label{f:3500k_fitted_lines}
\end{figure*}

\end{appendix}

\end{document}